\documentclass[useAMS,usenatbib,usegraphicx]{mn2e}
\usepackage{times}

\title[Understanding the MW spiral structure using the local kinematic groups]{Understanding the spiral structure of the Milky Way \\using the local kinematic groups}
\author[Antoja et al.]{T. Antoja,$^{1,2}$\thanks{E-mail: antoja@astro.rug.nl}  
F. Figueras,$^2$ M. Romero-G\'omez,$^2$ 
 \newauthor B. Pichardo,$^3$ O.Valenzuela$^3$ and E. Moreno$^3$  \\  
$^1$Kapteyn Astronomical Institute, University of Groningen, PO Box 800, 9700 AV Groningen, the Netherlands\\
$^2$Dept. d'Astronomia i Meteorologia, Institut de Ci\`encies del Cosmos (ICC), Universitat de Barcelona (IEEC-UB),\\ Mart\'i Franqu\`es 1, E08028 Barcelona, Spain\\
$^3$Instituto de Astronom\'ia, Universidad Nacional Aut\'onoma de M\'exico, A.P. 70-264, 04510, M\'exico, D.F., M\'exico\\
}

\date{Released 2002 Xxxxx XX}

\pagerange{\pageref{firstpage}--\pageref{lastpage}} \pubyear{2011}

\newcommand{\kms}{\rm\,km\,s^{-1}}
\newcommand{\kmskpc}{{\rm\,km\,s^{-1}{kpc}^{-1}}}
\newcommand{\pc}{\rm\,pc}
\newcommand{\kpc}{{\rm\,kpc}}
\newcommand{\Myr}{{\rm\,Myr}}
\newcommand{\Gyr}{{\rm\,Gyr}}
\newcommand{\degg}{{^\circ}}
\def\sun{{_\odot}}
\newcommand{\Rsun}{{R_\odot}}
\newcommand{\UVplane}{$U$--$V$ plane }
\newcommand{\UVplanef}{$U$--$V$ plane}
\newcommand{\UVplanepl}{$U$--$V$ planes}
\newcommand{\name}{{\tt PERLAS }}
\newcommand{\namef}{{\tt PERLAS}}
\newcommand{\namebf}{{\bfseries\ttfamily  PERLAS} }
\newcommand{\Msun}{{\mbox{M$_\odot$}}}
\def\mnras{MNRAS}
\def\aap{A\&A}
\def\apj{ApJ}
\def\aj{AJ}
\def\aaps{A\&AS}
\def\rmxaa{Rev. Mexicana Astron. Astrofis.}

\begin{document}

\maketitle

\label{firstpage}

\begin{abstract}

We study the spiral arm influence on the solar neighbourhood stellar kinematics. As the nature of the Milky Way (MW) spiral arms is not completely determined, we study two models: the Tight-Winding Approximation (TWA) model, which represents a local approximation, and a model with self-consistent material arms named \namef.  
This is a mass distribution with more abrupt gravitational forces. We perform test particle simulations after tuning the two models to the observational range for the MW spiral arm properties. We explore the effects of the arm properties and find that a significant region of the allowed parameter space favours the appearance of kinematic groups. 
The velocity distribution is mostly sensitive to the relative spiral arm phase and pattern speed. 
In all cases the arms induce strong kinematic imprints for pattern speeds around $17\kmskpc$ (close to the 4:1 inner resonance) but no substructure is induced close to corotation. 
The groups change significantly if one moves only $\sim0.6\kpc$ in galactocentric radius, but $\sim2\kpc$ in azimuth. 
The appearance time of each group is different, ranging from 0 to more than 1 $\Gyr$. 
Recent spiral arms can produce strong kinematic structures. 
The stellar response to the two potential models is significantly different near the Sun, both in density and kinematics. 
The \name model triggers more substructure for a larger range of pattern speed values. 
The kinematic groups can be used to reduce the current uncertainty about the MW spiral structure and to test whether this follows the TWA. 
However, groups such as the observed ones in the solar vicinity can be reproduced by different parameter combinations. 
Data from velocity distributions at larger distances are needed for a definitive constraint.
\end{abstract}


\section{Introduction}\label{introduction}

The spiral arms of our Galaxy have been typically characterised by radio observations of the 21-cm line of neutral hydrogen, giant HII regions, CO emission, and optical data of young O and B stars (e.g. \citealt{oort58,simonson70,georgelin76}). \citet{drimmel01} used FIR and NIR emission from the COBE/DIRBE to fit a model for the spiral arms in the stellar component of the Galaxy. Recently, \citet{benjamin05} and \citet{churchwell09}, based on infrared data from the Spitzer/GLIMPSE survey, and \citet{reid09}, using trigonometric parallaxes and proper motions of masers in high-mass star-forming regions, reported new results about the spiral structure of the Milky Way (MW). Despite the effort, there are still several caveats in the spiral arm properties in our Galaxy, such as pattern speed, strength, orientation and even the number of arms or their stellar or gaseous structure. Furthermore, the nature of the spiral arms themselves, i.e. their origin or their lifetime, are nowadays a matter of debate (\citealt{sellwood10}).

Apart from direct methods to detect spiral arm over-densities, an alternative method is based on the analysis of kinematic groups that are induced by the spiral arms in the local velocity distribution. Some of the moving groups in the solar neighbourhood were originally thought to be remnants of disrupted disc stellar clusters. However, the age or metallicity distribution of their stars are in contradiction with this hypothesis \citep{bensby07,antoja08}. \citet{kalnajs91} suggested that a moving group could be a group of stars that crosses the solar neighbourhood following a certain type of orbit induced by the Galactic bar gravitational potential. The demonstration that the bar resonances can produce a kinematic group, similar to the observed Hercules group \citep{dehnen00,fux01}, has lead to the exploration of similar effects on the velocity distribution due to the spiral arms (see \citealt{antoja10} for a review). This type of kinematic structures could depend strongly on some characteristics of the bar and spiral arms and, therefore, they result useful for our understanding of the MW large-scale structure and dynamics in its present and past form.

Few studies, however, have focused on the spiral arm effects on the disc velocity distribution. By integration of test particle orbits, \citet{desimone04} showed that few, but intense stochastic spiral density waves produce structures which are arranged in branches, resembling those from observations \citep{skuljan99,antoja08}. \citet{quillen05} developed a method to quantify, through orbital integration, the velocity distribution that would result from the existence of spiral-induced families of periodic orbits in the solar neighbourhood. They found, for models with the Sun in the outer limits of the 4:1 resonance and certain spiral arm orientation, two periodic orbits in the kinematic positions of Coma Berenices and Hyades--Pleiades. The test particle simulations by \citet{chakrabarty07} showed that the combined effect of a bar and 4 weak spiral arms, is necessary to reproduce the main local moving groups (Hercules, Hyades, Pleiades, Coma Berenices and Sirius). In the spiral-only models, more structures but less bold than those with the bar simulations were obtained. Finally, \citet{sellwood10a} showed that the angle-action variables of the stars in the Hyades stream could be consistent with the effects of a recent inner Lindblad resonance of a multi-arm (number of arms $m>2$) and transient pattern. 

All this work has shown that the spiral arms are able to induce kinematic groups in the local velocity distribution. However, there is no study that explores the influence of each spiral arm property using potential models designed according to the latest observational evidences for the MW spiral structure. On top of that, previous studies have modelled the spiral arm potential with a cosine function, following the Tight-Winding Approximation (TWA, e.g. \citealt{bible08}). However, it is not clear to us whether the MW spiral arms satisfy the conditions for self-consistency of this model, in particular, regarding their tightness and weakness. Moreover, the TWA implies that the spiral arms are a steady global mode of the disc but, as mentioned, it is still a matter of discussion whether the spiral arms are long-lived or transient structures.

Motivated by the recent picture of the MW spiral arms and the possibility that the TWA is not suitable for our Galaxy, we study the spiral arm effects on the solar vicinity kinematics with the TWA but also a different model, namely the \name model \citep{pichardo03}. This is a 3D mass distribution from which more abrupt gravitational potential and the forces are derived. As proved in \citet{franco02}, this difference has far-reaching consequences on the gaseous dynamical behaviour. Differences in the stellar response are expected too. In \citet{antoja09} we presented a preliminary study of the kinematic effects of this model. Here we aim to determine the conditions that favour the appearance of kinematic groups such as the ones that we observe near the Sun. We explore the influence of the spiral properties and discuss to which extent we can use the kinematic imprints to constrain the spiral arm properties and nature. We also compare the local kinematic imprints of the \name model with the ones of the TWA to see whether these imprints are useful to test the spiral arm gravitational potential modelling. To do this, we first tune these two models to the latest observational determinations for the properties of the MW spiral arms. For some properties only a range of evidence is available in the literature. Then we perform test particle simulations with both models considering these ranges, different initial conditions and integration times. 

Section \ref{modelMW} deals with the observed properties of the MW spiral structure. In Section \ref{modeling} we elaborate on the gravitational potential modelling of the spiral arms. Section \ref{models} presents the two spiral arm models and we fit them to the MW spiral arms. In Section \ref{force} we contrast the force fields of the two models. Section \ref{SIM} describes the test particle simulations and initial conditions. Afterwards, in Section \ref{comparison0} we compare the response in density and kinematics to the two models. We explore the influence of spiral arm characteristics, such as pattern speed, orientation and integration time, on the local velocity distribution in Sections \ref{patternspeed}, \ref{maximum} and \ref{time}. In Section \ref{constrain} we discuss whether it is currently possible to constrain the spiral arm properties by using the local observed kinematic groups. Finally, we summarise our main results and their implications.

\section{The spiral structure of the MW}\label{modelMW}

Most of the properties of the MW spiral structure remain rather undetermined. Here we establish a range of plausibility for each property in order to tune our spiral arm models to the MW spiral arms. Table \ref{tab:MWarms} shows these adopted ranges. 

{\bf Locus.} The geometry of the MW spiral arms is still a matter of intense debate. Different estimates of the pitch angle and even the number of arms can be found from different tracers (see \citealt{vallee08}). Maps of OB-associations and HII-regions, CO emission, masers in high-mass star-forming regions (\citealt{georgelin76}, \citealt{taylor93}, \citealt{vallee08} and references therein, \citealt{reid09}) show a 4-armed pattern, usually referred as Sagittarius--Carina, Scutum--Centaurus (or Scutum--Crux), Perseus and the outer (or Cygnus) arms. However, according to COBE K-band observations \citep{drimmel01} and to the infrared Spitzer/GLIMPSE survey \citep{churchwell09}, only two of these four arms are major MW arms. These are traced by the stellar (young and mainly old) population as enhancements in the spiral tangencies. External galaxies often show different morphologies in blue and near-infrared colours (Grosb{\o}l, Patsis \& Pompei 2004). 
Although it might not be a general property of galaxies \citep{eskridge02}, many galaxies classified as flocculent or multi-armed systems in blue, display a 2-armed grand design spiral in the K band (\citealt{block94}, Kendall, Kennicutt \& Clarke 2011). 
 Also hydrodynamic models have shown that it is possible to form arms of compressed gas, without increasing the stellar surface density, as a response to a 2-armed pattern \citep{martos04}. Here we consider only these two stellar major arms ($m=2$) as they trace the underlying mass distribution of our Galaxy. 

\begin{figure}
\centering
\includegraphics[width=0.4\textwidth]{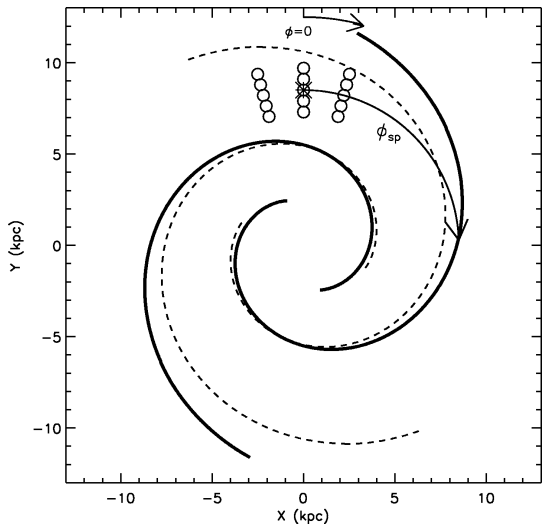} 
\caption{Adopted locus for our spiral arm models: locus 1 (solid line) and locus 2 (dashed line). The star at $X=0$ and $Y=8.5\kpc$ indicates the assumed solar position. Open circles indicate several regions near the solar neighbourhood.}  
\label{f:mw}
\end{figure}

The exact location of the stellar spiral arms is still unclear and part of the current models in the outer regions of the MW are an extrapolation of observations in the inner regions. \citet{benjamin05}, \citet{churchwell09} and \citet{vallee08} consider that the stellar arm counterpart of Scutum--Centaurus is the Perseus arm. But according to the model by \citet{drimmel01}, the counterpart goes far beyond the Perseus arm in the anti-centre direction. Because of these discrepancies, we will adopt two different locus: the one reported by \citet{drimmel01} (hereafter locus 1, solid black curve in Fig. \ref{f:mw}), and the fitting by \citet{vallee08} for the Scutum and Perseus arms (hereafter locus 2, dashed black curve in Fig. \ref{f:mw}). 

\begin{table}
 \caption{Assumed values or ranges for the properties of the MW spiral arms.}\label{tab:MWarms}
\centering
\begin{tabular}{llc}\hline
 \multicolumn{2}{l}{Property}                        &Value or range   \\   \hline
Number of arms          &$m$                           & $2$             \\
Scale length            &$R_\Sigma$ ($\kpc$)           & $2.5$           \\
Locus beginning         &$R_{sp}$ ($\kpc$)             & $2.6/3.6$           \\
Pitch angle             &$i$  ($\degg$)                & $15.5/12.8$     \\
Relative spiral phase   &$\phi_{sp}(R_\sun)$ ($\degg$) & $88/60$    \\
Pattern speed           &$\Omega_{sp}$ ({$\kmskpc$})   & 15--30         \\  
Density contrast        &$A_2$                         & 0.14--0.23      \\  
Density contrast        &$K$                           & 1.32--1.6      \\  \hline
 \end{tabular}
\end{table}

We use the galactocentric cylindrical coordinates ($R$, $\phi$) with the azimuth $\phi>0$ in the direction of rotation and origin as indicated in Fig. \ref{f:mw}. The locus of one arm is obtained through solving $2(\phi - {\phi}_0) + g(R) = 0$, with the condition $\phi \leq {\phi}_0$. The second arm is obtained by symmetry. The constant $\phi_0$ fixes the arm orientation and it corresponds to the azimuth of the line that joins the two starting points of the spiral locus. It is $-20\degg$ for locus 1 and $-70\degg$ for locus 2. The function $g(R)$ defines the spiral shape. We adopt the one from \citet{roberts79}:
\begin{equation}\label{e:locusdrimmel}
g(R)=\left( \frac{2}{N \tan i}\right) \ln\left( 1+\left(\frac{R}{R_{sp}} \right)^N\right),
\end{equation}
where $R_{sp}$ is the radius of the beginning of the spiral shape locus. The parameter $N$ measures how sharply the change from a bar-like to spiral-like occurs in the inner regions. Here the limit of $N\rightarrow\infty$ is taken, approximated by $N=100$, which produces spiral arms that begin forming an angle of $\sim90\degg$ with the line that joins the two starting points of the locus. We have checked that our simulations (which are studied in the outer disc) do not depend on the exact shape in these inner parts. 

The pitch angle is $i=15.5\degg$ and $12.8\degg$, for locus 1 and 2, as estimated by \citet{drimmel01} and \citet{vallee08}, respectively. The relative spiral phase $\phi_{sp}(R_\sun)$ is the azimuth between the Sun's position and the peak of the spiral at the same radius (curved long arrow in Fig. \ref{f:mw}). It is $88\degg$ and $60\degg$ for locus 1 and 2, respectively. The Sun's is initially assumed to be at $R=8.5\kpc$ and $\phi=0\degg$ (star in Fig. \ref{f:mw}). But we will examine also nearby regions (open circles in Fig. \ref{f:mw}) for which  the relative spiral phase varies approximately from $-45\degg$ to $45\degg$ for locus 1, and from $-85\degg$ to $10\degg$ for locus 2. With this we are more flexible in the spiral arm location which, as mentioned, is rather undetermined. 

{\bf Pattern speed.} Estimations for the spiral structure pattern speed $\Omega_{sp}$ come from open clusters birthplace analysis, kinematics of young stars, and comparisons of the observed $^{12}$CO (l,v) diagram with models for the gas flow (see \citealt{gerhard10} for a review). For the two-armed K-band model by \citet{drimmel01} used in the present study, a pattern speed of $20\kmskpc$ gave the best consistency results for the stellar response to the spiral pattern \citep{martos04}. But we will explore the range 15--30$\kmskpc$, assuming rigid rotation, as deduced from several literature determinations \citep{gerhard10}. 
This range gives a rotation period between 400 and 200 $\Myr$. 

{\bf Density contrast.} Determinations for the density contrast of the MW spiral structure are few and entail a large uncertainty. Moreover, the density contrast definition is sometimes ambiguous. For external galaxies the spiral amplitude is often quantified as $A_2$ \citep{grosbol04}, which is the amplitude of the $m=2$ component of the Fourier decomposition of the surface brightness scaled to azimuthally averaged surface brightness. A different measure is the arm--interarm contrast $K(R)=(1+A_2)/(1-A_2)$ \citep{bible08}. If $A_2$ is measured from infrared bands, a mass to light ratio of $\sim1$ can be assumed \citep{kent92} and these quantities are directly a measure of the density contrast $A_2\sim\delta\sigma/\sigma_0$ (or $K=(\sigma_0+\delta\sigma)/(\sigma_0-\delta\sigma)$) where $\sigma_0$ is the axisymmetric surface density, and $\delta\sigma$ is the enhancement of density on the spiral arms. 

Table \ref{tab:densitycontrast} summarises several determinations of the density contrast in the literature. \citet{drimmel01} give an arm--interarm ratio in the K band surface brightness in our Galaxy of $K=1.32$ (or, equivalently, $A_2=0.14$). However, these authors report that this may be undervalued due to underestimation of the arm scale height as compared to that of the disc. Indeed, this contrast is significantly smaller than in external galaxies, where values up to $A_2=0.6$ are found \citep{rix95}. Recent data from the GLIMPSE survey give an excess of 20--30\% 
of stellar counts at the maximum, with respect to the axisymmetric exponential fitting \citep{benjamin05}. For the exponential fitting, these authors have subtracted the enhancements in the spiral arms. Therefore, we ascertain that their value is a ratio with respect to the minimum stellar counts. Although the conversion to density contrast is not straightforward, we will assume that this gives directly $K\sim1.3$ ($A_2\sim0.13$), which is not far from the \citet{drimmel01} value. We will take $A_2=0.14$ as a lower limit for the MW from \citet{drimmel01}. 

\begin{table}
 \caption{Density contrast of the MW spiral arms from several studies. }\label{tab:densitycontrast}
\begin{tabular}{lcc}\hline
Author&$K$ &$A_2$\\\hline
\citet{drimmel01} & 1.32  & 0.14  \\
\citet{benjamin05}& 1.30  & 0.13  \\
\citet{grosbol04} & 1.2--1.6  & 0.1--0.23  \\\hline
\end{tabular}
\end{table}

Another determination for the MW spiral density contrast can be deduced from the relation between pitch angle and spiral amplitude for external galaxies, explored in \citet{grosbol04}. For a pitch angle of $15.5\degg$, the contrast $A_2$ ranged approximately from 0.1 and 0.23. Galaxies with a contrast up to 0.5 were also found, but the authors explain that this may be overestimated due to the strong star formation in these galaxies. Therefore, we will assume an upper limit of the density contrast of $A_2=0.23$.

Instead of being measured globally or as a function of radius as in external galaxies, for our own Galaxy density contrast estimations come from data of certain positions of the disc (mainly spiral tangencies, i.e. at a given radius). In the absence of detailed information about the density contrast along the MW arms, we assume that the chosen density contrast range is for an intermediate radius of $R\sim6\kpc$. Besides, we adopt an exponential fall as in \citet{contopoulos86} with scale length of $R_\Sigma=2.5 \kpc$. 

\section{Modelling the spiral arm gravitational potential: two different approaches}\label{modeling}

Now we deal with the models for the gravitational potential of the arms. One approximation to model spiral arms would be the one given by \citet{gerola78}, who suggested that self-propagating star formation in a differentially rotating disc is capable of producing large scale spiral features. This description produces stochastic spiral arms and, although this may be able to explain flocculent spiral arms, it would neither be the case for about the half of galaxies that present grand design spirals in the infrared bands \citep{kendall11} nor the case of the MW where clear K-band arms are observed (Section \ref{modelMW}). Therefore, we do not consider a gravitational model for the flocculent arm type.

The first model that we consider in this study is the TWA spiral arms. This density wave theory (Lin, Yuan \& Shu 1969) 
 has extensively been used in stellar and gas dynamics simulations. From its birth, this spiral structure theory has been capable to provide certain qualitative explanation, going from some local kinematics of our own Galaxy to systematic changes in spiral arm properties along the Hubble sequence. However, it has never been straightforward to obtain a good fitting between density wave theory, and observations, especially, of course, when galaxies reach a non-linear regime, which might be the case in the majority of galaxies. One simple example among several: the density wave theory says that stronger shocks of material (gas, stars) with spiral arms should be observed in thinner spiral arms, but it seems that wider arms produce stronger shocks \citep{kennicutt82}.

The spiral arms obtained through the TWA are an elegant solution to stability analysis of galactic discs, by perturbing the basic equations of stellar dynamics \citep{lin69}. Treated as weak perturbations (low mass, low pitch angles) to the background potential, a simple cosine expression for the spiral arm potential is a self-consistent solution for the linear regime \citep{lin69}. Nevertheless, it has been a costume to extrapolate this mathematical approximation from the linear domain to all kind of spiral galaxies, under the general assumption that spiral arms are weak perturbations in galaxies. A key question then is whether the TWA can be applied to the MW spiral arms. First, the TWA solution is limited to tightly wound spiral arms with $m/\tan i>>1$. As stated in \citet{bible08}, this is satisfied in most galaxies but not with a comfortable margin. According to the observational constraints for our Galaxy (Section \ref{modelMW}), pitch angles of $i=15.5\degg$ (locus 1) and $12.8\degg$ (locus 2) give $7.2\leq m/\tan i \leq 8.8$, and, therefore, the assumption is at least doubtfully satisfied. Second, the TWA is a perturbative solution for small density contrasts. Although the determinations for the MW spiral density contrast entail large uncertainty, the maximum spiral arm density contrast of $23\%$ of the axisymmetric disc seems to exceed the requiment. Due to the uncertainity on the MW spiral arm characteristics, it is not clear yet whether they can be compatible with the self-consistent TWA solution.

The second studied model for the spiral arm gravitational potential is the \name model (sPiral arms potEntial foRmed by obLAte Spheroids, \citealt{pichardo03}). Unlike the cosine potential of the TWA, which represents a local approximation to the spiral arms, the nature of \name is a very different one. It is a model with material arms in the sense that it corresponds to a given spiral arm mass distribution, from which the potential and forces are derived. In particular, the model is constructed as a superposition of small pieces of mass distribution. Because of this, it is more flexible than the TWA. It can be easily adjusted to the MW spiral arms or any spiral density profile that might be far from following a cosine function, as the profiles in some external galaxies \citep{kendall11}. For this model, we can adjust the total spiral mass, the arm width and arm height, which is not possible for the TWA model. This potential results more realistic in the sense that it considers the force exerted by the entire spiral arm structure, sculpting much more complicated shapes in potential and force than a simple cosine function. These intrinsic differences may induce significant deviations on orbital dynamics from the classic cosine. In addition, it is a full 3D model which, instead of taking an ad hoc dependence on the $z$ coordinate (e.g an additional potential term ${\rm sech} ^2[z/z_s]$), it considers directly a three-dimensional mass distribution. It also satisfies a periodic orbit diagnostic for self-consistency \citep{pichardo03}, which consists in analysing the stellar orbital reinforcement of the potential as in \citet{patsis91}. As for the TWA, the orbital self-consistence of the model is only assured for a certain range of parameters, especially mass and pattern speed. Next we present in detail these two potentials and how they are tuned to the recent observations of the MW spiral structure.

\section{The models}\label{models}

The axisymmetric part of our MW models is taken from \citet{allen91} (hereafter A\&S). It is composed by a bulge, a flattened disc and a massive spherical halo. The first two are modelled as Miyamoto--Nagai potentials \citep{miyamoto75} and the halo is built as a spherical potential. The main adopted observational constraints of the model are summarised in table 1 of A\&S. A value of $\Rsun=8.5\kpc$ for the Sun's galactocentric distance and a circular speed of $V_c(R\sun)=220\kms$ are adopted. The total axisymmetric mass is $M_T=9\times 10^{11} M_{\odot}$. The local circular frequency is $\Omega(R\sun)=25.8\kmskpc$. The two spiral models,  TWA and \namef, are described in next sections.

\subsection{The TWA spiral arms}\label{TWAdes}

In the TWA the spiral arm potential is obtained by solving the Poisson's equation for a plane wave in a razor-thin disc \citep{bible08,lin69}, after modelling the spiral structure as a periodic perturbation term to the axisymmetric disc. Following \citet{contopoulos86}, the TWA gives a potential in the plane of the form:
\begin{equation}
\Phi_{sp} (R,\phi)=-A_{sp}Re^{-R/R_\Sigma}\cos\bigl(m[\phi-\phi_0]+g(R)\bigr),
\end{equation}
where $A_{sp}$ is a measure of the amplitude of the spiral pattern, and the rest have already been defined. The amplitude of the spiral perturbing potential, following again \citet{contopoulos86}, it is related to the amplitude of the perturbed surface density (through Poisson's equation) as:
\begin{equation}\label{eq:contrastTWA}
\delta\sigma(R)=\frac{A_{sp}e^{-R/R_\Sigma}}{\pi G |\tan i|}.
\end{equation}

\begin{figure}
\centering
\includegraphics[width=0.35\textwidth]{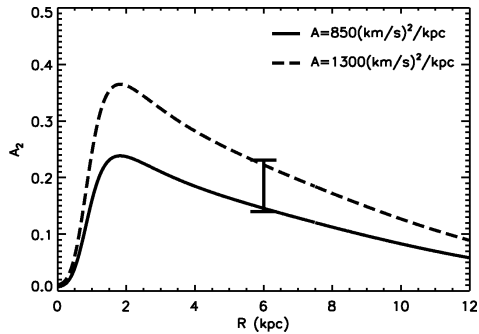}
\caption{Density contrast $A_2=\delta\sigma/\sigma_0$ as a function of radius of the TWA model for the two amplitudes that fit the high and low observational limits (vertical error bar) for the MW.} \label{f:contrastTWA}
\end{figure}

\begin{table}
 \caption{Assumed parameters of the TWA model.}\label{tab:arms10}
\centering
\begin{tabular}{llc}\hline
\multicolumn{2}{l}{Parameter} &Range\\   \hline
Amplitude (locus 1)      &$A_{sp}$ ($[\kms]^2\kpc^{-1}$) &850--1300 \\
Amplitude  (locus 2)     &$A_{sp}$ ($[\kms]^2\kpc^{-1}$) &650--1100 \\ \hline
 \end{tabular}
\end{table}

To fit this model to the MW spiral arms we use the locus and parameters as discussed in Section \ref{modelMW}. In particular for this model, the density contrast $A_2=\delta\sigma(R)/\sigma_0(R)$ is obtained directly from Equation (\ref{eq:contrastTWA}) and the surface density $\sigma_0$ of the stellar part of the axisymmetric A\&S model (disc and bulge). Due to the different dependence with $R$ of $\delta\sigma$ and the disc surface density in A\&S, the density contrast $A_2$ decreases with radius. As explained in Section \ref{modelMW}, we chose the range of $A_{sp}$ that fits the observational limits for the density contrast $A_2$ at a intermediate radius of $R\sim6\kpc$. In Table \ref{tab:arms10} we show the determined range. For locus 1, this is $A_{sp}=$850--1300 $[\kms]^2\kpc^{-1}$. The density contrast as a function of radius for these two limits is shown in Fig. \ref{f:contrastTWA}. For locus 2 (with a smaller pitch angle) the range is $A_{sp}=$650--1100 $[\kms]^2\kpc^{-1}$. 

Notice also that Equation (\ref{e:locusdrimmel}) gives a locus that joins the two starting points of the spiral locus in the disc central part (not plotted in Fig. \ref{e:locusdrimmel}). However, we have checked that our simulations (which are studied in more outer regions) do not depend on the exact shape in these inner parts.

\subsection{The \namebf spiral arms}\label{PERLASdes}

The spiral arms of the \name model consist of a mass distribution which is built as a superposition of inhomogeneous oblate spheroids along a given locus. A linear density fall, inside each spheroid, is considered with zero density at its boundary. The central density of the spheroids follow an exponential fall in $R$ along the arms with scale length of Table \ref{tab:MWarms}. The potential and force fields for these spheroids are given in \citet{schmidt56}. The overlapping of spheroids allows a smooth distribution along the locus, resulting in a continuous function for the gravitational force. We checked that no significant change was observed if this separation was decreased thus increasing the smoothness of the spiral mass distribution. For more details about the construction of the model see \citet{pichardo03}.

\begin{figure}
\centering
\includegraphics[width=0.35\textwidth]{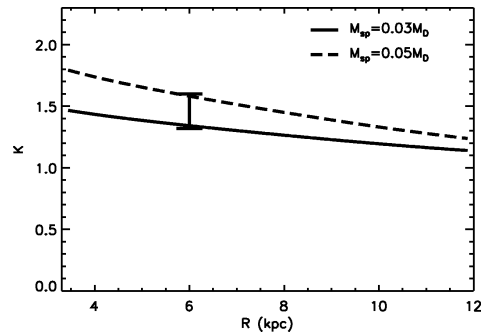}
\caption{Density contrast $K=(\sigma_0+\delta\sigma)/(\sigma_0)-\delta\sigma$ as a function of radius of the \name model for the two amplitudes that fit the high and low observational limits (vertical error bar) for the MW.} 
\label{f:contrastPERLAS}
\end{figure}

\begin{table}
 \caption{Assumed parameters of the \name model.}\label{tab:arms11}
\centering
\begin{tabular}{llc}\hline
 \multicolumn{2}{l}{Parameter} &Values or range\\   \hline
Beginning of the arms        &$R_i$ ($\kpc$)                 &$3.3$                    \\
End of the arms              &$R_f$  ($\kpc$)                &$12$                     \\
Arm half-width               &$a_0$  ($\kpc$)                &$1$                      \\
Arm height                   &$c_0$ ($\kpc$)                 & $0.5$                   \\
Mass (locus 1)               &$M_{sp}/ M_D$                  & 0.03--0.05     \\
Mass  (locus 2)              &$M_{sp}/ M_D$                  & 0.035--0.06     \\\hline
 \end{tabular}
\end{table}

To fit this model to the MW spiral arms we use the locus and parameters as discussed in Section \ref{modelMW}. Additional parameters for this model are shown in Table \ref{tab:arms11}. In this case, the spiral amplitude is quantified through the value $M_{sp}/ M_D$ which is the ratio of the spiral mass to the mass of the disc in the A\&S model. In the \name model the spiral arms are added as an mass enhancement to the axisymmetric background on the imposed locus.\footnote{Note the difference with the TWA for which the arms consist of a density perturbation that adds density in the arm region but subtracts density in the interarm region.} Because of this, for the \name model the mass of the spiral arms ($M_{sp}$) is globally subtracted from the original disc of the A\&S model ($M_D$), which guarantees that the total mass of the model do not change and neither does the mean circular velocity. For this reason, the parameter $K$ is more suitable to be related with the $M_{sp}/ M_D$, as it is the ratio of the surface density on the spiral arm (axisymmetric disc and spheroids) to the minimum density (axisymmetric disc). As for the TWA model, the density contrast $K$ decreases with radius. For locus 1, the range of spiral mass ratio that fits the observational range of density contrast $K$ at $R\sim6\kpc$ is $M_{sp}/ M_D=$0.03--0.05 (Fig. \ref{f:contrastPERLAS}). For locus 2, the respective range is\footnote{Contrary to the the TWA model where a smaller pitch angle demanded a lower amplitude to reproduce a given density contrast, for the \name model a higher mass ratio is needed. This is because we fix the spiral arm end at $R=12\kpc$ and for locus 2 with a more tightly wound arm, the arm longitude is larger than for locus 1. This implies that a higher total spiral mass is necessary to reproduce the desired contrast.} $M_{sp}/M_D=$0.035--0.06

Additionally, the beginning of the spheroid superposition, i.e. the effective arm beginning, is at $3.3\kpc$ and $3.6\kpc$, respectively for locus 1 and locus 2. The superposition ends at $12\kpc$ for both cases. The arm half-width and height above the plane are taken to be $1\kpc$ and $0.5\kpc$, respectively, according to the analysis of a sample of external galaxies \citep{kennicutt82} and the discussion in \citet{martos98}.

\section{Force fields of the spiral models}\label{force}

The main difference between the \name spiral arms and the TWA falls on the construction itself. In the TWA, the spiral arms are a small perturbative term of the potential. This produces a cancellation of the contribution from the distant parts of the pattern to the local force. The \name spiral arms correspond to an independent mass distribution. The contribution from the entire spiral pattern causes the spiral potential and force to adopt shapes that are not correctly fit by the simple TWA perturbing term that has been traditionally employed. In this section we show these fundamental differences in detail by studying the force field exerted by these two models using the Fourier decomposition. If not stated the contrary, the plots and values in this section are referred to the lower limit of the density contrast defined in Section \ref{modelMW}.

\begin{figure}
\centering\includegraphics[width=0.4\textwidth]{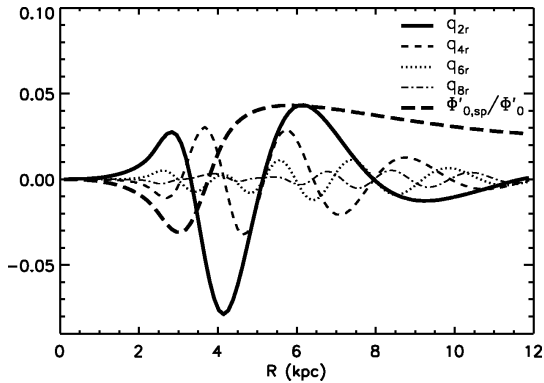}
\caption{Decomposition of the radial force of the \name model in the terms $q_{mr}$ for $m=2,4,6,8$ of Equation (\ref{eq:qr}). Long dashed line is the ratio of the radial axisymmetric part of the force of \name model to the total axisymmetric radial force $\Phi'_{0,sp}/\Phi'_0$.} \label{f:decomp}
\end{figure}

First we study the parameter $q_{mr}$, which is the $m$ term of Fourier decomposition of the non-axisymmetric radial force scaled to the axisymmetric radial force. This is:
\begin{equation}\label{eq:qr}
q_{mr}(R)=\frac{\Phi_m'(R)}{\Phi_0'(R)}\, ,
\end{equation}
where primes denote derivatives with respect to $R$, and $\Phi_m(R)$ is the $m$-component of the Fourier decomposition of the global potential, that is of the potential of the spiral arms plus the  A\&S potential. In particular for $m=0$, $\Phi_0'(R)$ corresponds to the axisymmetric part of the global potential. The TWA model consists of a pure and simple $m=2$ component. Therefore, the spiral arms only contribute in the numerator of Equation (\ref{eq:qr}), and the axisymmetric model A\&S is the denominator. By contrast, the \name arms consist of a more complex potential structure with more than a $m=2$ term. In Fig. \ref{f:decomp} we show the Fourier decomposition of the \name potential. In particular, the long dashed line shows the ratio of the axisymmetric component of the radial force of the \name model $\Phi'_{0,sp}$ to the axisymmetric global radial force $\Phi'_0$. We see that the axisymmetric radial force of the \name model can be as large as 4\% of the total axisymmetric force (and up to 7\% for high limit of the density contrast). This important $m=0$ component contributes, toguether with the A\&S model, to the axisymmetric part of the global potential (denominator in Equation (\ref{eq:qr})) together with the A\&S model. In the same plot we see that for this model the $m=2$ component is dominant but it also has non-vanished $m>2$ terms $q_{mr}$.

\begin{figure}
\centering
\includegraphics[width=0.4\textwidth]{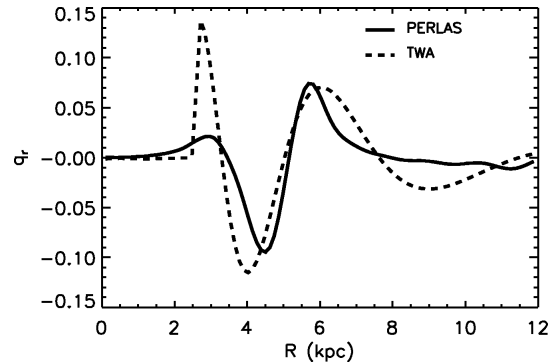}
\caption{Radial spiral force $q_{r}$ as a function of radius for the TWA and the \name model.} \label{f:qrsp}
\end{figure}

In order to quantify the whole non-axisymmetric force of our spiral arm models we define the parameter:
\begin{equation}\label{eq:qr2}
q_r(R)=\sum_{m\geq2} q_{mr}(R)
\end{equation}
which is the ratio of all the non-axisymmetric radial force terms to the axisymmetric part of the global potential.\footnote{Note that often, e.g. in \citet{athanassoula83}, the strength of the non-axisymmetric components is quantified through the parameter $q_{r}$ which, contrary to the present study, only includes the $m=2$ component.}
 For the TWA, $q_r=q_{2r}$. But for the \name model it is the contribution of all terms $q_{mr}$, with $m\geq2$, of Fig. \ref{f:decomp}. Fig. \ref{f:qrsp} shows $q_r$ as a function of radius for the two models. The range of the radial force amplitude of the two models is approximately the same as they are fitted to reproduce the same density contrast. However, the detailed shape of $q_{r}(R)$ is different due to the important contribution of all terms $m\geq2$ in the \name model. From this figure we also observe that the maximum radial force of \name spiral arms with respect to the total axisymmetric radial force is 0.09 (absolute value). Around $R\sun$ the parameter $q_{r}$ is $0.003$. For the TWA, the maximum is 0.14 and the value at solar radius is 0.03. For the higher limit of the density contrast range, the shape of $q_{r}(R)$ does not change. But in that case, the \name model produces a maximum radial force with respect to the axisymmetric background of 0.16 and a value near the solar radius of 0.005. For the TWA the maximum value is 0.21 and the value at the solar radius is 0.04.

\begin{figure}
\centering\includegraphics[width=0.4\textwidth]{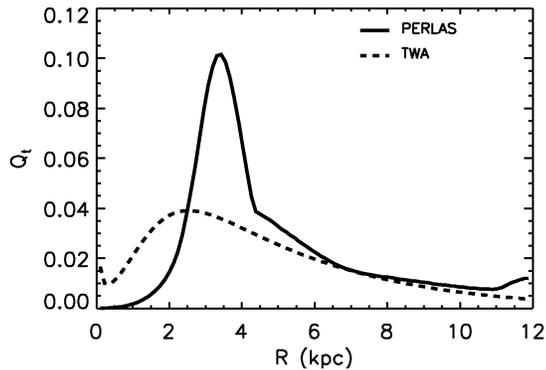}
\caption{Maximum tangential spiral force $Q_T$ as a function of radius for the TWA and the \name model.} \label{f:Qtsp}
\end{figure}

The parameter $Q_T(R)$ was used in \citet{combes81} to quantify the tangential force of bars. Here we use it for spiral arms as in \citet{block04}. It is the ratio of the maximum spiral arm tangential force at a given radius to the mean total radial axisymmetric force at that radius:
\begin{equation}\label{eq:QT}
Q_{T}(R)=\frac{F_\phi^{max}}{<F_{R0}(R)>}\, .
\end{equation}
The parameter $Q_T$ includes the tangential forces due to all terms $m\geq2$ for the \name model. This parameter is shown in Fig. \ref{f:Qtsp}. The large and small bumps at $R\sim3.5\kpc$ and $12\kpc$ for the black curve are due to the more abrupt beginning and end of the arms in the \name model but have no consequences for our study that is focused on near solar radius. The amplitude range of $Q_t$ is rather similar in the radii of interest in this study (7--10$\kpc$). The maximum of $Q_T(R)$ gives a single and quantitative measure of the torque or strength of the spiral arms \citep{block04} and is called $Q_{s}$. For the \name model we find that $Q_s=0.1$ and for the TWA it is 0.039. For the maximum density contrast, $Q_s$  is 0.17 for the \name model and 0.060 for the TWA. In \citet{block04} the spiral strengths $Q_s$ in a sample of 15 external galaxies range from 0 to 0.46. According to this, the models for the MW spiral arms are below the median of external galaxies. For both models $Q_T(R_\sun)$ is $\sim0.01$ for the lower density contrast limit and $\sim0.02$ for the high case.

\begin{figure}
\centering
\includegraphics[width=0.45\textwidth]{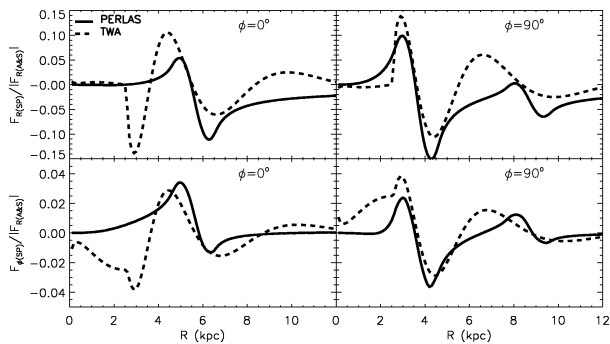}
\caption{Radial force (top) and tangential force (bottom) as a function of radius for two different azimuths $\phi$ for the TWA and \name model. The forces are scaled to the A\&S radial axisymmetric force.} \label{f:frspiral}
\end{figure}

\begin{figure}
\centering
\includegraphics[width=0.45\textwidth]{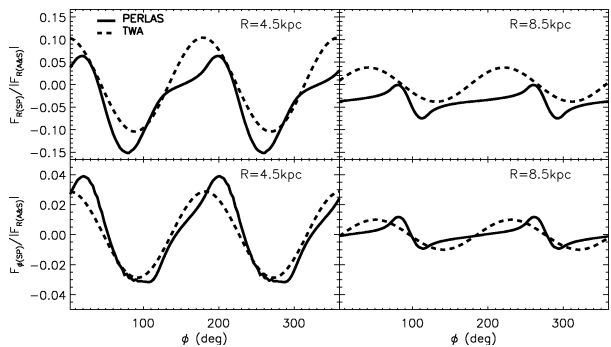}
\caption{Same as Fig. \ref{f:frspiral}, but as a function of azimuth for two different radius $R$.} \label{f:faspiral}
\end{figure}

The parameters $q_r$ and $Q_T$ measure average and maximum values at a given radius, respectively. We now study the detailed force profile. Figs. \ref{f:frspiral} and \ref{f:faspiral} show the radial and tangential forces of the two models as a function of azimuth $\phi$ and radius $R$ scaled to the axisymmetric radial force of the A\&S model. We see significantly different force fields. In general, the \name spiral arms (solid line) presents more abrupt features than the TWA. It also has different locations ($R$, $\phi$) for the minima and maxima for both the radial and tangential forces. The positions at which the force changes its sign is different as well. Besides, the radial force of the \name model is in general below the force for the TWA, that is non-symmetric with respect to 0 and shifted to negative values. This is due to the inner enclosed spiral mass at each radius, that is the $m=0$ component of the \name spiral force mentioned before. For instance, in the top right panel in Fig. \ref{f:faspiral} (solar radius $R=8.5\kpc$) the radial force is negative for all $\phi$, whereas the TWA consists of a symmetric oscillation around 0 force. 

To conclude, it is quite different to model the gravitational potential by using the TWA (i.e. a cosine function, that fits a given locus and certain density contrast) or by building a mass distribution on the same locus and contrast. We have seen that the two approaches to model the spiral arms gravitational potential give significantly different force profile. In next sections we study how these differences propagate to the kinematic stellar response.


\section{The simulations}\label{SIM}

In order to study the effect of the spiral arms on the kinematic distribution near the Sun, we perform numerical integrations of test particle orbits, as most of the studies up to now. These are simple models and their self-consistency is not assured (Section \ref{modeling}). By contrast, models such as N-body simulations of galaxy formation could model self-consistently the spiral arm kinematic effects, as well as include naturally the bar or spiral arm evolution and other processes that might also sculpt the velocity distribution (e.g. past accretion events, star formation bursts). However, N-body simulations with larger than current number of disc particles, better spatial and temporal resolution, and models similar to the MW are required. At this moment, test particle simulations allow us to use models that fit the MW spiral arms in a controlled manner. They are simpler and faster models that offer easy exploration and understanding of the influence of the spiral arm properties. 

For our simulations we adopt the potential models of Section \ref{models} and the initial conditions described in Section \ref {ic}. In Section \ref{procedure} we give more details about the method. Our study focuses on the kinematic effects produced on or near the plane. For this we assume that the vertical motion is decoupled with the in-plane movement. This is reasonably true for nearly circular orbits that do not take larger height above the plane \citep{bible08}. As our initial conditions consist of rather cold discs (see Section \ref{ic}), we adopt these assumptions and simplify our analysis by considering 2D simulations ($z=0$). 

 The $(U,V)$ velocity reference system is used which is centred on a given position on the disc plane and moves following the Regional Standard of Rest (RSR). This is defined as the point located at a radius $R$ that describes a circular orbit around the Galactic Centre with a constant circular speed $V_c(R)$. $U$ is the radial velocity component, which is positive towards the Galactic Centre, and $V$ is the tangential component, positive in the direction of Galactic rotation.

\subsection{The initial conditions}\label{ic}

We explore two different types of initial conditions: IC1 \& IC2. Both are axisymmetric discs, following an exponential profile with scale length of $R_\Sigma=2.5 \kpc$. This is similar to the thin disc scale length found by \citet{freudenreich98} from COBE/DIRBE data. 
IC1--  The velocity distribution relative to the RSR is adopted as a Gaussian with low dispersions $\sigma_U=\sigma_V=5\kms$ and constant for all radii. These values are similar to the induced gas velocity dispersions in the plane due to Galactic spiral shocks \citep{kim09} and to the dispersion of the youngest Hipparcos stars \citep{aumer09}. With this IC we aim to simulate a cold young disc. 

IC2--  For these IC, the phase space distribution function (DF) is constructed, as discussed in \citet{hernquist93}. The velocity field is approximated using the moments of the collisionless Boltzmann equation, simplified by the epicyclic approximation. Following \citet{dehnen99}, we will adopt $\sigma_U^3$ proportional to the surface density, and therefore, the radial velocity dispersion profile is: \begin{equation}\label{e:sigmaU}
\sigma_U(R)=\sigma_{0}e^{-R/(3R_{\Sigma})}.
\end{equation}
The local normalisation is chosen $\sigma_U(R_\odot)\sim20 \kms$. The tangential velocity dispersion profile is derived from the epicyclic approximation 
(equation 4.317 of \citealt{bible08}).
Also we take into account the asymmetric drift 
 (equation 4.228 of \citealt{bible08}). Assuming that the disc is stationary, axisymmetric and symmetric about its equator, and that the orientation of the velocity ellipsoid is aligned with the coordinate axes, it is: 
\begin{equation}\label{e:ad}
v_a=\displaystyle\frac{\sigma_U^2}{2V_c}\left[\displaystyle\frac{-B}{A-B}-1+R\left(\frac{1}{R_\Sigma}+\frac{2}{3R_\Sigma}\right)\right].
\end{equation}
This DF confers to the IC2 disc the properties of an intermediate population ($\sim 1\Gyr$) of the MW thin disc \citep{holmberg09}.

\subsection{Integration procedure and analysis}\label{procedure}

In our simulations the integration of each particle is initialised at a time $t=-\tau$ (as a convention) and ends at t$=0$, being $\tau$ the particle exposure time to the potential. The time $\tau$ is chosen at random between $0$ and $2\Gyr$ for each particle. This maximum integration time corresponds to $\sim7$ revolutions of the spiral arms for a typical pattern speed of $20\kmskpc$. With this procedure, the final velocity distributions result of a superposition of particles integrated different times, resembling the observed distribution, with a superposition of stars of different ages. In some cases, we separate the particles into different bins of integration times, to study the induced kinematic effects as a function of time.

The integration of the motion equations is done with the Bulirsch--Stoer algorithm of \citet{press92}, conserving Jacobi's integral within a relative variation of $|(E_{Ji}-E_{Jf})/E_{Ji}|\approx 10^{-9}$. The reference frame used for the calculations is the rotation frame of the spiral arms. The number of particles in each simulation is about $10^7$. 

After the integration, we study the induced kinematic distribution near the Sun. For the analysis, we focus on the \UVplane of the 15 circular regions indicated in Fig. \ref{f:mw}, which have $300\pc$ of radius (similar to the radius of the available observed local velocity distribution, see fig. 1 in \citealt{antoja08}). The centres of the regions are located at five different radius (7.3, 7.9, 8.5, 9.1 and $9.7\kpc$) separated $600\pc$ from the nearest ones, and at azimuths $-15\degg$, $0\degg$ and $15\degg$, which is found to be the approximate azimuth interval that shows significant differences between adjacent \UVplanepl. The symmetries of the Galactic potential ($\Phi(R,\phi)$=$\Phi(R,180\degg+\phi)$) allow us to double the number of particles in each studied region. In all cases, the number of particles in the final velocity distributions is larger than $\sim10000$, being statistically robust. 
Fixed the spiral arm pattern speed, these 15 regions have different ratio $\Omega_{sp}/\Omega$, that is, the pattern speed scaled to the local azimuthal frequency of that region. These are representative of different pattern speeds, allowing us to explore small variations in this parameter. Also they have different relative spiral phase $\phi_{sp}$ which aims to explore the range of  uncertainty in this parameter, as explained in Section \ref{modelMW}. 

We apply the Wavelet Denoising method (WD) to the velocity  distributions. This method gives a smooth DF from a discrete point distribution via a smoothing/filtering treatment at different scales, that eliminates Poisson fluctuations \citep[for details, see][]{antoja08}. The results will be shown with logarithmic colour scale and contours representing, from inside out, 0.01, 0.05, 0.10, 0.20, 0.30, 0.40, 0.50, 0.6, 0.70, 0.8, 0.9 and 0.999 of the maximum density.

\section{Stellar response to the TWA and \namebf models}\label{comparison0}

We have run simulations for pattern speeds $\Omega_{sp}$ of 15, 18, 20, 22, 25 and $30\kmskpc$ (inside the observational range of evidence, Section \ref{modelMW}) for both models. We have used the two limits of density contrast, the two loci and the different initial conditions detailed in previous sections. Here we compare the results for the two models. First, in Section \ref{response} we analyse the density response to the spiral models in our simulations. With this we aim to study the consistency between density response and gravitational potential of the models, and to relate the density response to the kinematic analysis of subsequent sections. 

\subsection{Density response}\label{response}

\begin{figure*}
\centering
\includegraphics[width=0.37\textwidth]{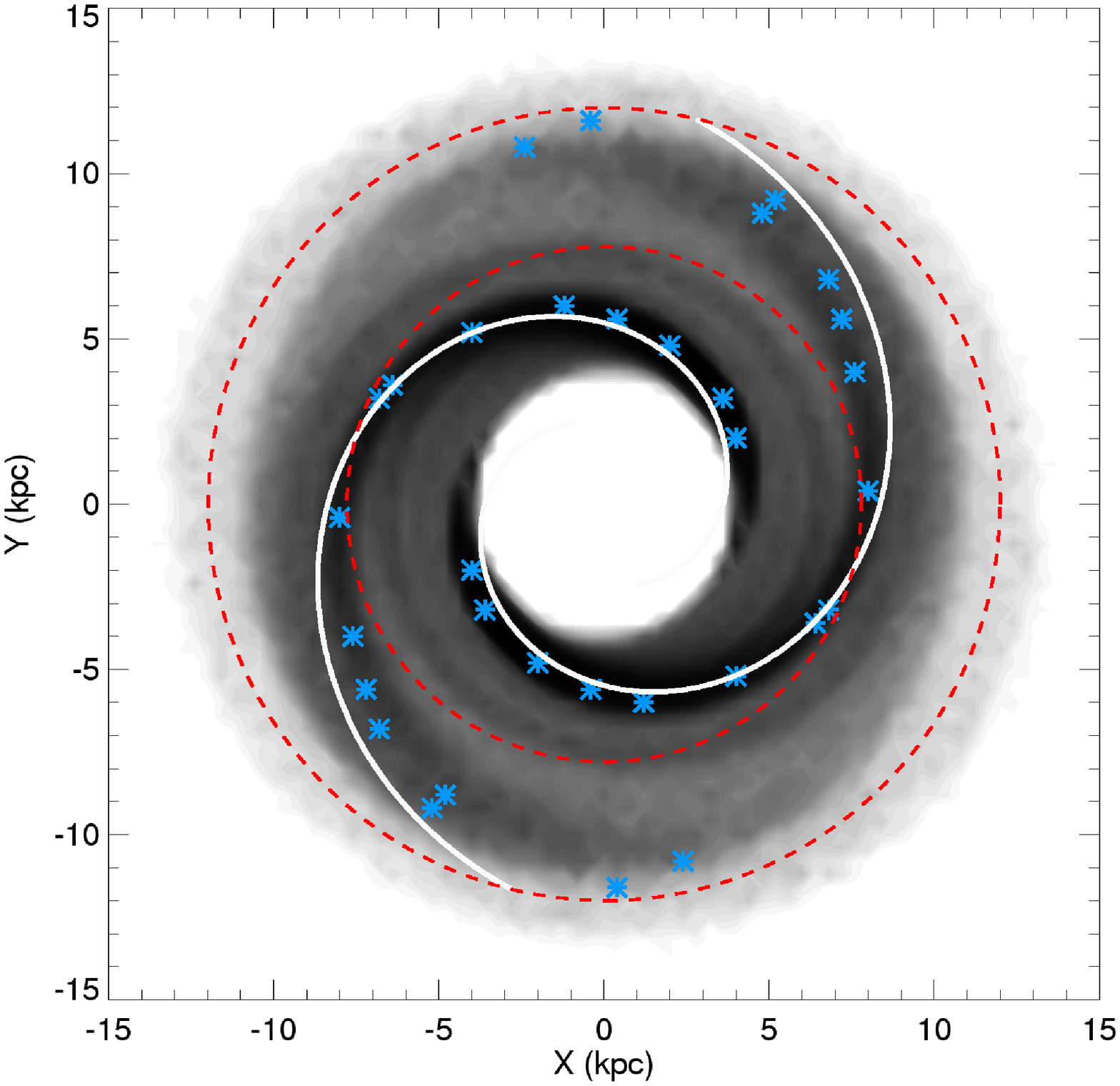}
\includegraphics[width=0.37\textwidth]{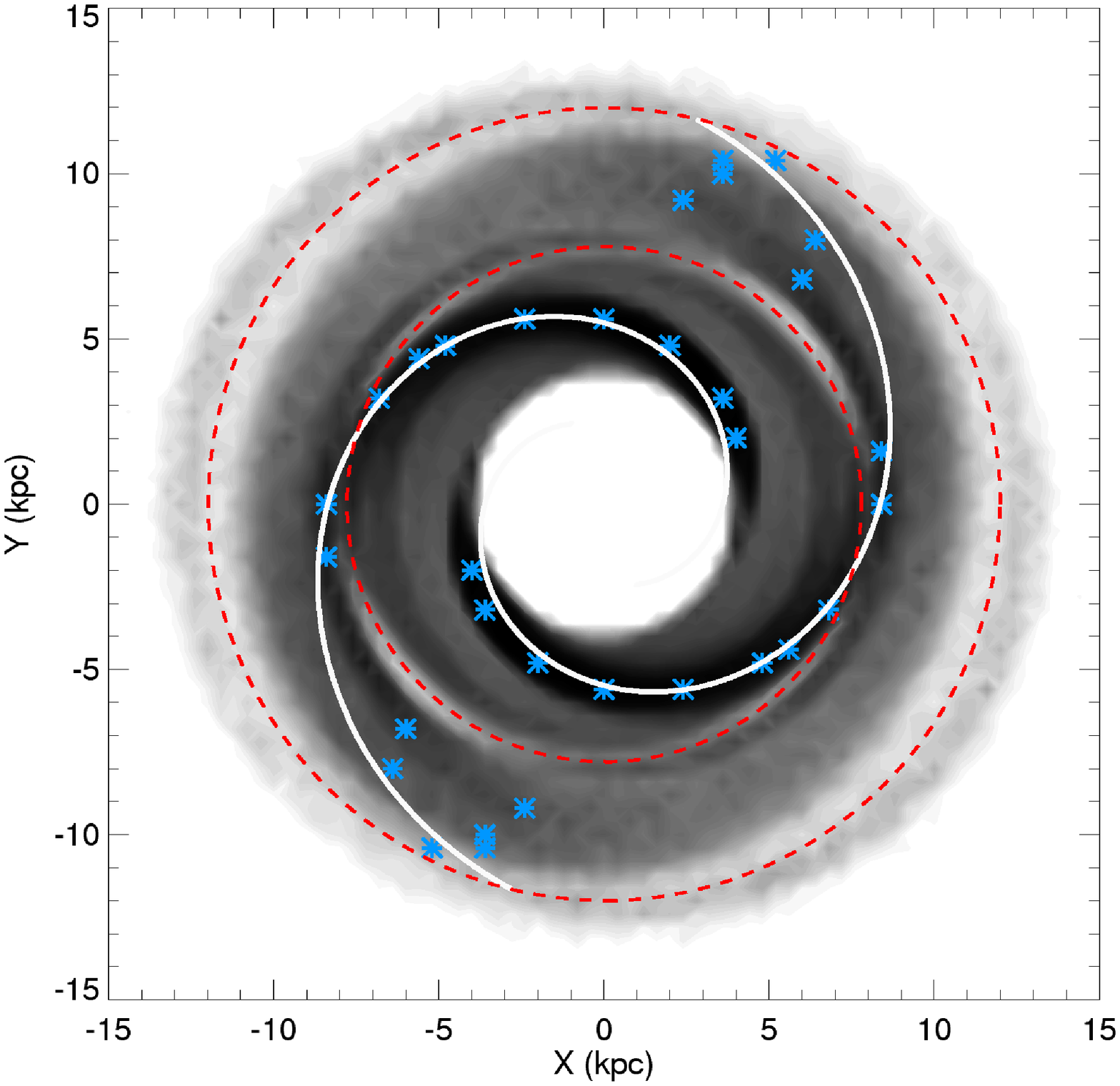}

\includegraphics[width=0.37\textwidth]{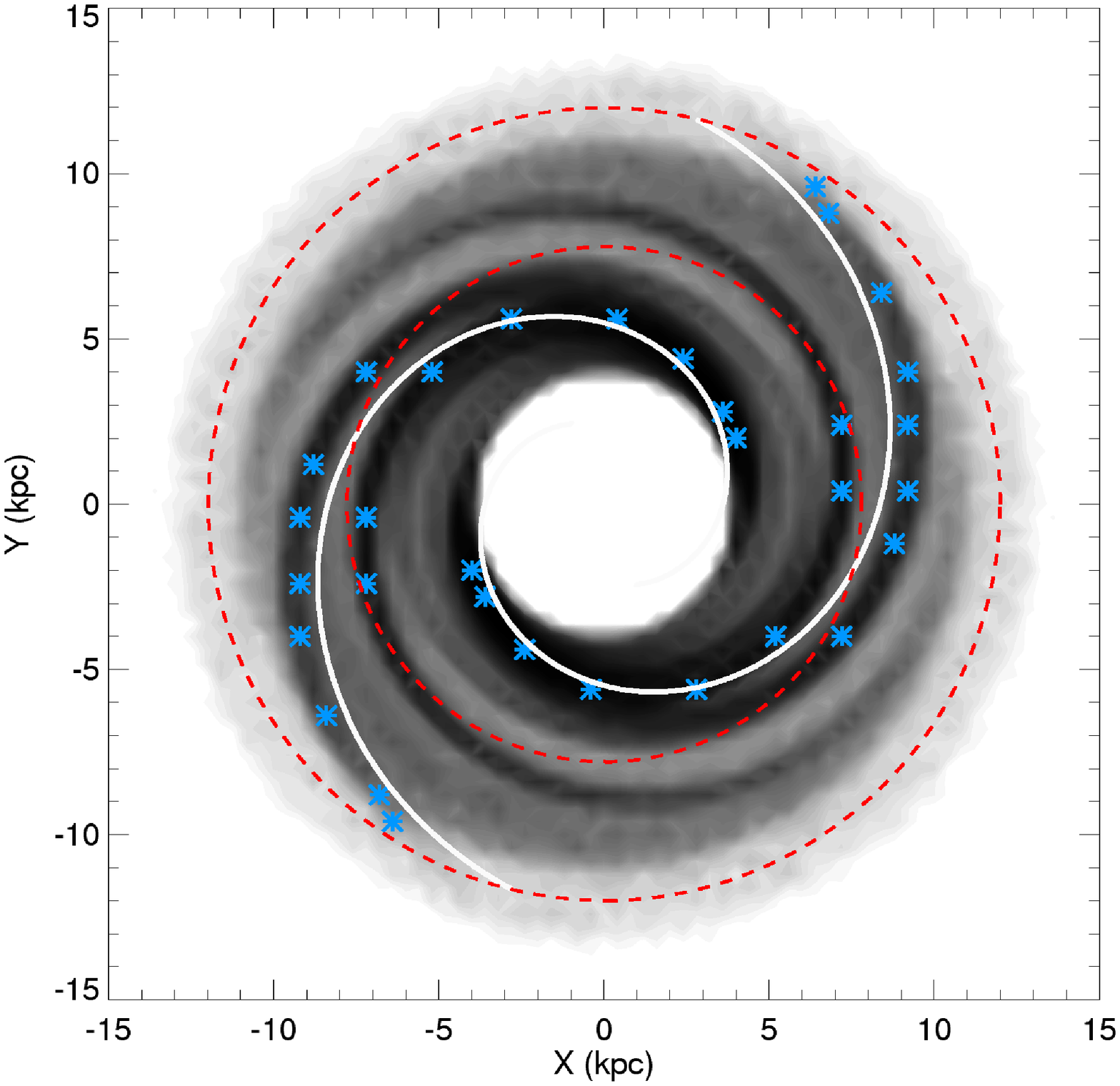}
\includegraphics[width=0.37\textwidth]{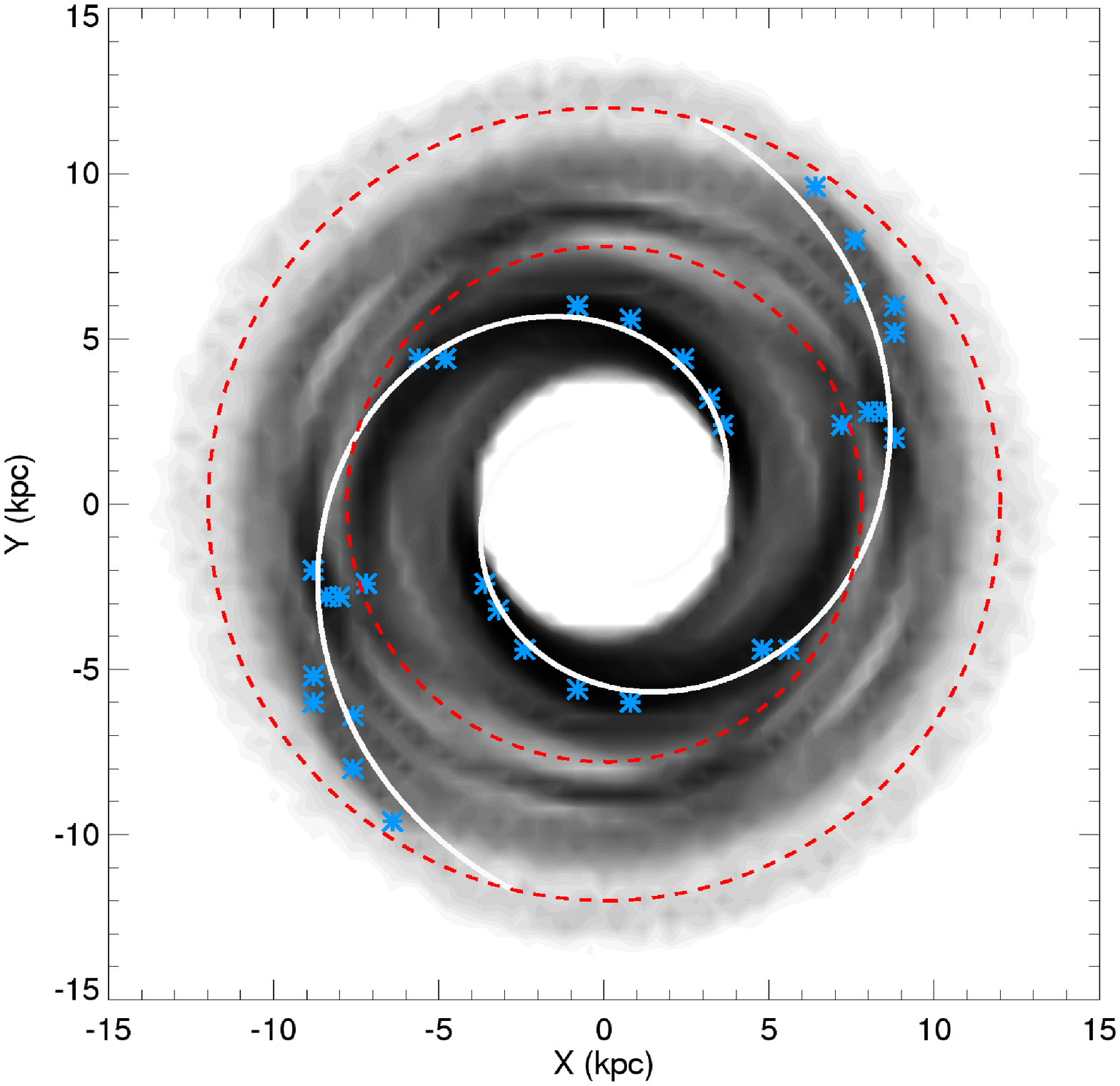}
\caption{Comparison between the imposed spiral locus and the density response to the TWA model (top) and \name model (bottom) for $\Omega_{sp}=18\kmskpc$, locus 1, IC1 and minimum density contrast. The density is scaled to the axisymmetric exponential density at the beginning of the simulation. First column shows the simulation for small integration times (0--400 $\Myr$) and second column corresponds to large integration times (1600--2000 $\Myr$). Red dashed circles show the approximate positions of the 4:1 inner resonances (inner circle) and corotation (outer circle), estimated from the axisymmetric model of A\&S.}  
\label{f:response}
\end{figure*}

Fig. \ref{f:response} shows the density distribution of a simulation with the TWA (top) and the \name model (bottom) taking the same conditions (pattern speed, locus, etc.) for both models. Darker regions show the denser regions with the same scale in all panels of the figure. 
 First and second column correspond to short and long integration time, respectively. Blue stars show the positions of maximum density in rings of $500\pc$
of radius. The white curves indicate the locus of spiral arm model. Inner and outer red dashed circles show the approximate positions of the 4:1 inner resonances and corotation, respectively.

First row shows that the density response to the TWA with the minimum density contrast and a pitch angle of $i=15.5\degg$ does not follow the spiral locus imposed in the potential approximately beyond the 4:1 inner resonance. The spiral arms in this simulation show a more tightly wound pattern at outer radius for short integration time and high dispersion for long integration times. Right panel also shows complex structures like rings and low density regions. The different pattern speeds and amplitudes in the ranges established for the MW give equivalent results.

\citet{contopoulos86} already studied the density response for several spiral models based on the TWA, mainly focusing on periodic orbits. They concluded that strong spirals can not exist beyond the 4:1 inner resonance as in this disc region the spiral response is out of phase due to non-linear effects. By contrast, weak or tightly wound spirals can exist up to corotation, as the linear theory predicts \citep{lin69}. Our results are in agreement with the results of \citet{contopoulos86} for strong spirals.\footnote{Preliminary tests seem to indicate that the non-linear effects in our simulations are due mainly to MW spiral arms that are not enough tightly wound. A simulation with the same pitch angle, but smaller amplitude, still presents the inconsistency between imposed pattern and response. On the contrary, a simulation with the same amplitude but smaller pitch angle follows almost perfectly the locus up to corotation for all integration times.} 
 An exhaustive analysis of the response density is out of the scope of this paper. But the difficulties in using the TWA tuned to the MW as a self-consistent model beyond the 4:1 inner resonance, poses important constraints in the use of the TWA for the MW. For certain pattern speeds, this would imply a too short ending radius of the spiral arms (e.g. the spiral arms could end at $\sim4.3\kpc$ for a pattern speed of $30\kmskpc$).\footnote{Notice that the existence of an additional slower spiral mode at outer radius could solve this problem. We are not considering this type of composite models in the present paper, but see discussion in Section \ref{bar}.} To conclude, the TWA might not be a proper description for the MW spiral arms for some parameter combinations (pattern speed, pitch angle, amplitude). 
 
The density response to the \name model (second row of Fig. \ref{response}) shows also disrupting effects near the inner 4:1 resonance but, at larger radius the density response does follow the imposed spiral arms for small and large integration times. Indeed, with larger pattern speeds, the response follow the imposed locus even beyond corotation.\footnote{It is worth mentioning that other models different from the TWA \citep{toomre81,rautiainen99,voglis06,romerogomez07} have shown that spiral arms can extend well up to or beyond corotation as it happens with the \name model.} Nevertheless, the orbital self-consistence of the \name model is only assured for a certain range of parameters, specially in mass and pattern speed (see \citet{pichardo03}). 

\begin{figure}
\centering
\includegraphics[width=0.4\textwidth]{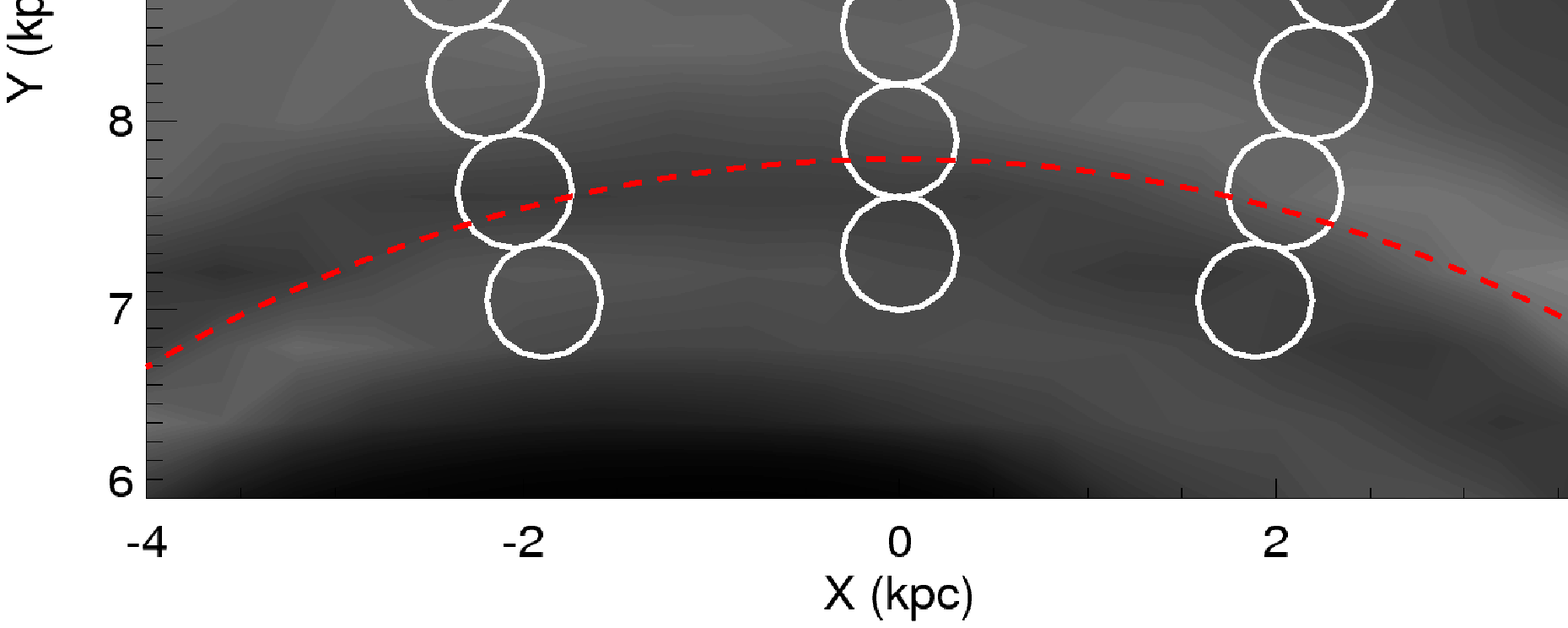}
\includegraphics[width=0.4\textwidth]{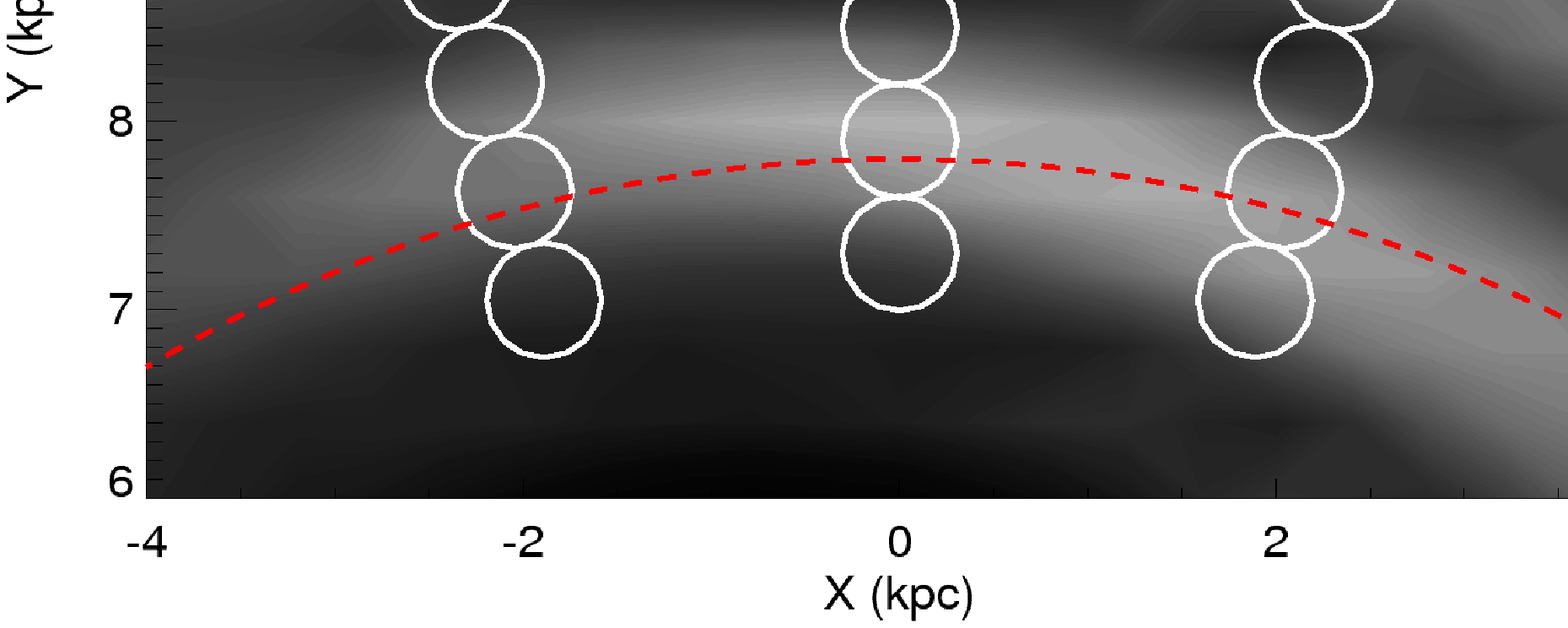} 
\caption{Zoom of the density response at solar position for the TWA model (top) and \name model (bottom) for all particles (all integration times) in the simulation of Fig. \ref{f:response}.}   
\label{f:densitylocal}
\end{figure}

For the \name model we also see complex density structures such as two extra arms for short integration times or low density regions and rings in the long integration case. All these patterns that are observed for both models could be related to the different resonances of the spiral arms and might well represent rings and spiral arm spurs that are seen in our Galaxy and in external galaxies \citep{elmegreen90}. For instance, two emptier regions are located on the 4:1 inner resonance at $R\sim8.0\kpc$ in the right panels. Notice, however, that these appear at different azimuths depending on the model (near $\phi=0\degg$ for the \name model but at an azimuth of $\phi\sim50\degg$ for the TWA). In fact, note that the induced density in our region of study (near solar positions) is significantly different between models. Fig. \ref{f:densitylocal} is the local density response for all particles of the simulation. The TWA model (top) presents softer density gradients, whereas the \name model (bottom) shows the mentioned emptier region at solar azimuth. The local density response of the two models is different in all our examined cases, emphasizing the distinct force fields studied in Section \ref{force} and forecasting the differences in the induced velocity distribution near solar positions (Section \ref{comparison}). 

Simulations with IC2 show qualitatively equivalent results to the ones with IC1 for both models, except for the fact that the density features appear less sharp but more diluted, as expected for a hotter and, therefore, less responsive disc. For both models we observe a transient nature of the density patterns. In all cases the density response becomes stationary approximately between 400 $\Myr$ and 1200 $\Myr$ of integration time depending e.g. on the pattern speed (except for outer radius $R>10\kpc$, out of the region of study). Similar time dependence will be observed for the kinematic distributions of these models (Section \ref{time}).

\subsection{Different kinematic imprints}\label{comparison}

\begin{figure}
\centering
\includegraphics[width=0.35\textheight]{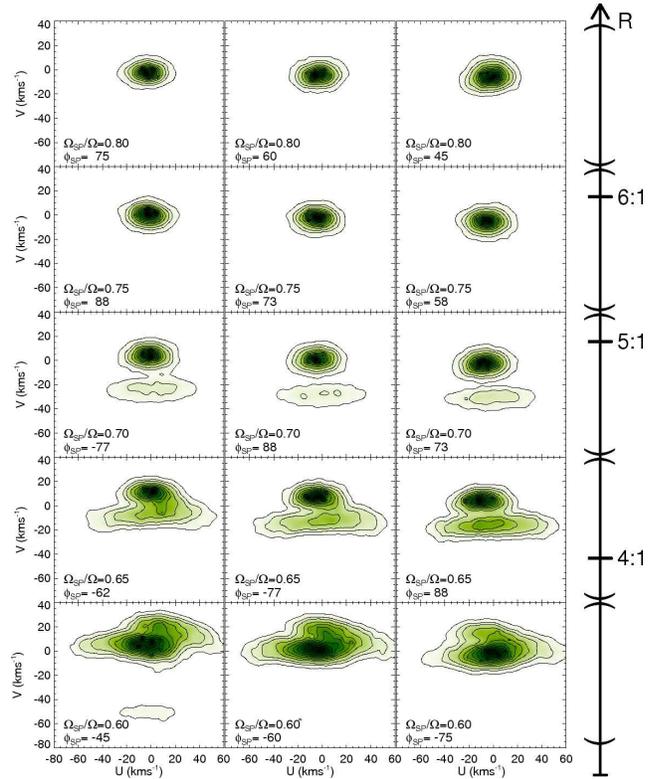} 
\caption{$U$--$V$ velocity distributions for the 15 regions of Fig. \ref{f:mw} for the simulations with the TWA model with locus 1, maximum density contrast, IC1 and a pattern speed of $18\kmskpc$. The 15 panels are positioned in the figure similarly to their spatial location of Fig. \ref{f:mw}. The scale in the right  indicates the spatial position of the spiral resonances (see text).}  
\label{f:18TWA}
\end{figure}

\begin{figure}
\centering
\includegraphics[width=0.35\textheight]{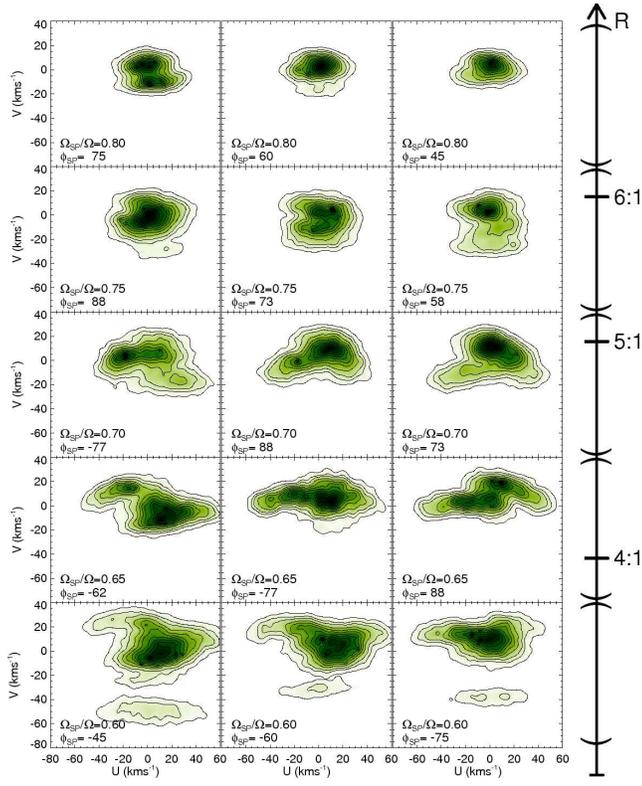} 
\caption{$U$--$V$ velocity distributions for the 15 regions of Fig. \ref{f:mw} for the simulations with the \name model with locus 1, maximum density contrast, IC1 and a pattern speed of $18\kmskpc$.}  
\label{f:speed18}
\end{figure}

In this section we compare the stellar kinematic response to the \name model and the TWA. As an example, we now adopt locus 1, maximum density contrast, IC1 and a pattern speed of $18\kmskpc$ for both cases. The \UVplane for the regions of Fig. \ref{f:mw} are shown in Figs. \ref{f:18TWA} and \ref{f:speed18} for the TWA and \name models, respectively. The 15 panels of each figure are positioned similarly to their spatial location of Fig. \ref{f:mw}. Each panel corresponds to a region with different relative spiral phase $\phi_{sp}$ and ratio $\Omega_{sp}/\Omega$, which are indicated in each panel. 

First we see that both models induce several and rich kinematic substructures near solar position. Nevertheless, we see that the details and induced groups at each of the 15 regions are different for each model, in agreement with the differences between induced density (Section \ref{response}). For instance, an elongated group at low $V$ appears for the left panel of the last row for the TWA model but it is present in all panels in this row (several spiral phases) for the \name model. The central part of these panels is also fairly different. The rest of the panels also show groups at different positions depending on the model. 

We have detected that the two models occasionally give a similar \UVplane when comparing different disc positions with comparable density distribution. For instance, we see similar velocity distributions near the emptier region on the 4:1 inner resonance (Section \ref{response}), which is is located at different azimuths depending on the model. We attribute this occasional shift to the differences in the force fields of both models (Section \ref{force}). We have seen that maxima, minima and 0 points of the force profiles (both tangential and radial) are locally shifted in radius and azimuth when comparing the two models. For example, in Fig. \ref{f:faspiral} we see that the maximum tangential force of the \name model is shifted to larger azimuths compared to the TWA. On the contrary, the minima are shifted to smaller azimuths. The link between phase space, force profile and resonance locations involves a large-scale study of the models which is out of the purpose of the present paper, but is currently under study (Antoja et al., in preparation). 

As common aspects of the kinematic response of the two models, we notice variety of shapes, sizes, inclinations and positions of the induced substructures, for this and other pattern speed ratios. For example, notice more or less rounded groups but also thin elongated structures. These and other simulations have shown that both models induce structures at $V$ as high as $30\kms$ and as low as $-60\kms$. The $U$ component of the induced kinematic planes can range from -80 to $60\kms$. The number of induced groups depends on the region but is in most cases 2 or 3. Interestingly, the kinematic groups are found to be in general more symmetric with respect to the $U=0$ axis for the TWA than in the \name model. 

More important, by inspecting these kinematic plots, we see that the \name model induces kinematic substructure where the TWA model does not (e.g. first and second rows). The TWA gives substructure for a smaller range of pattern speeds. We examine this in detail in Section \ref{maximum}. In the following sections, we study the influence of the spiral arm properties on the kinematic distribution near the solar position. We focus basically on the \name model but we contrast the results with the ones obtained by the TWA model when important differences arise.

\section{Effects of the pattern speed and spiral arm orientation}\label{patternspeed}

\begin{figure}
\centering
\includegraphics[width=0.35\textheight]{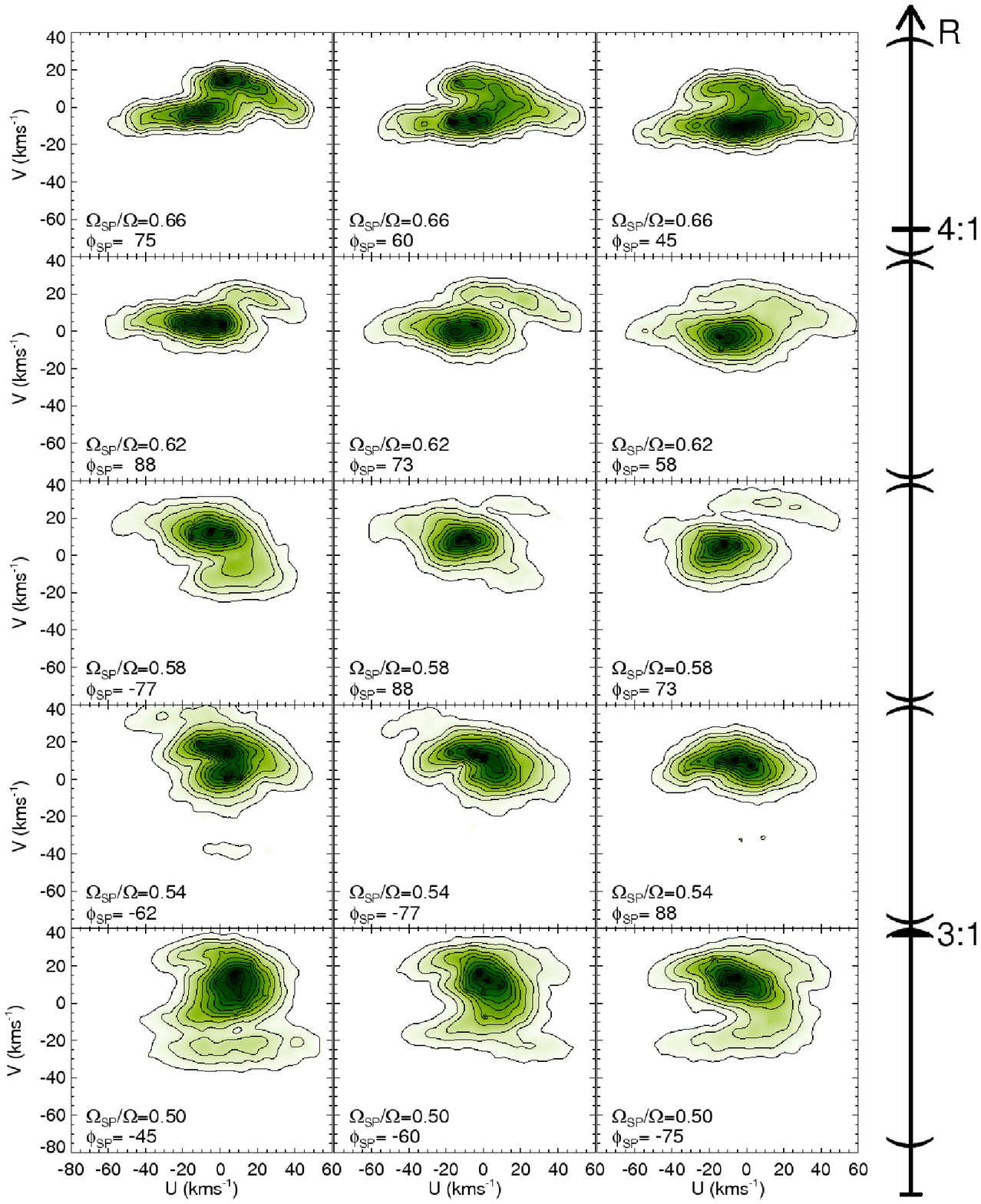} 
\caption{Same as Fig. \ref{f:speed18} but for a spiral pattern speed of $15\kmskpc$.}  
\label{f:speed15}
\end{figure}

\begin{figure}
\centering
\includegraphics[width=0.35\textheight]{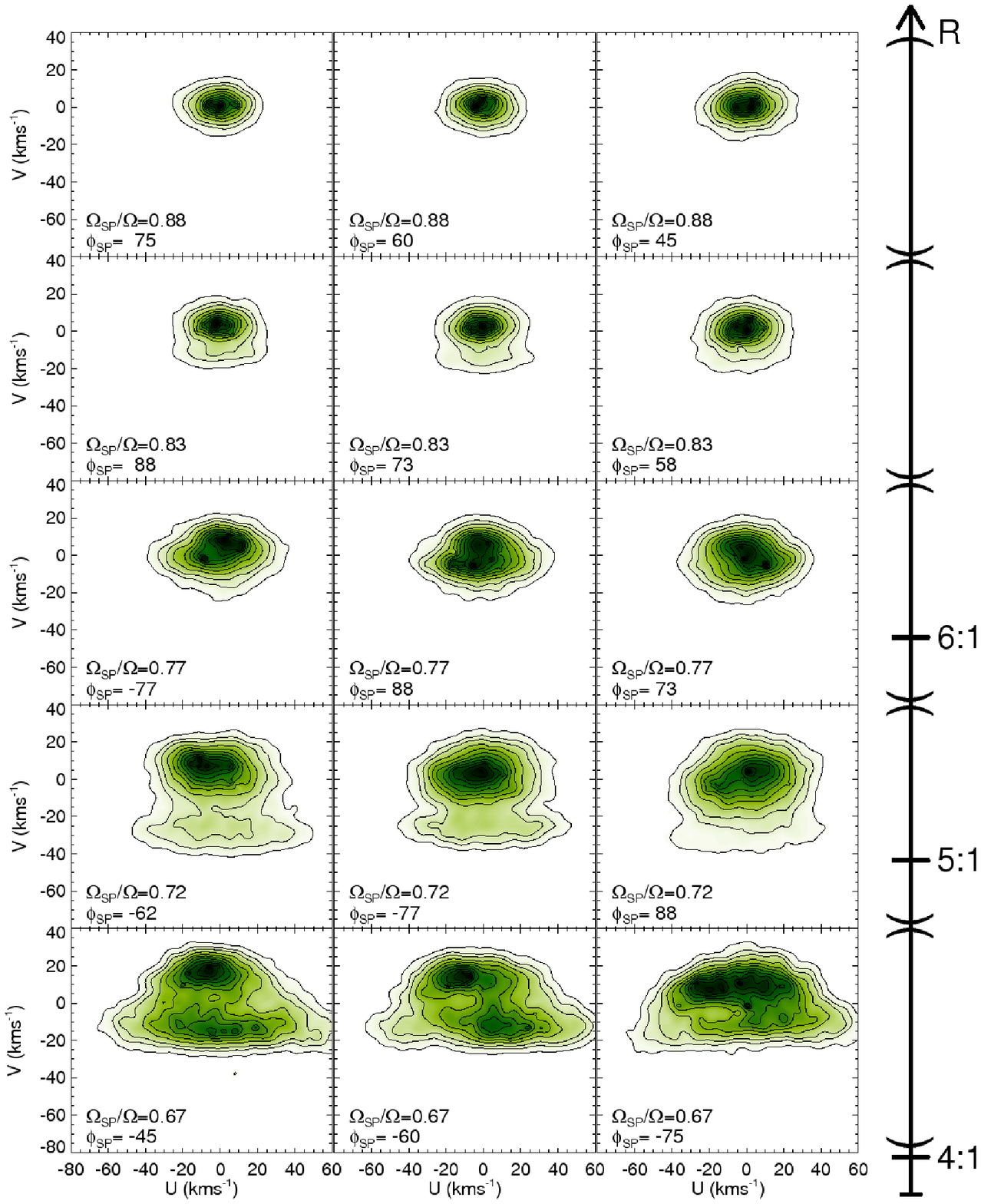} 
\caption{Same as Fig. \ref{f:speed18} but for a spiral pattern speed of $20\kmskpc$.}  
\label{f:speed20}
\end{figure}

Figs. \ref{f:speed15} and \ref{f:speed20} show the contour plots of the \UVplane for the 15 regions shown in Fig. \ref{f:mw} of the simulations corresponding the pattern speeds of $15\kmskpc$ and $20\kmskpc$, respectively, for the \name model. Together with the already presented Fig. \ref{f:speed18}, these panels allows us to study the effects of several pattern speed ratios $\Omega_{sp}/\Omega$ and relative spiral phases $\phi_{sp}$. In all cases we use locus 1, maximum density contrast and IC1. 

The panels show rich kinematic substructure near the solar position, strongly depending on the pattern speed ratio and relative phase of the spiral arm. The kinematic substructures change more slowly with azimuth (along rows) than with radius (along columns). The changes are significant if one moves only $\sim0.6\kpc$ in radius. But $\sim 2\kpc$ are needed in azimuth to detect important differences in the \UVplanef. The kinematic groups slightly change position on the \UVplane with azimuth or radius. For instance, for a pattern speed of $\Omega_{sp}/\Omega\sim 0.65$ (fourth row in Fig. \ref{f:speed18}) the velocity distributions show basically two groups, elongated in the $U$ direction, which appear in different positions in the $U$ component depending on the azimuth.

There are equivalent \UVplane plots in different simulations. For instance, top left panel of Fig. \ref{f:speed15} and third panel of the fourth row of Fig. \ref{f:speed18} correspond to similar relative pattern speed $\Omega_{sp}/\Omega$ and spiral phase $\phi_{sp}$. However, moving in radius is not strictly equivalent to changing the pattern speed. First the spiral strength or density contrast depends on radius (see Section \ref{models}). Second, although the initial velocity dispersion of particles in IC1 is flat with radius, at the end of the simulation this increases for inner radius, as we have seen with a simulation only including the axisymmetric part of the model. One example is the group at $V\sim-40\kms$ in the right panel in the last row Fig. \ref{f:speed18} which is not present in the equivalent panel of Fig. \ref{f:speed15} (left panel of the second row). The smaller strength of the spiral at outer radius and/or the smaller velocity dispersion do not allow to crowd this structure for the simulation in Fig. \ref{f:speed15}. 

\begin{figure}
\centering
\includegraphics[width=0.35\textheight]{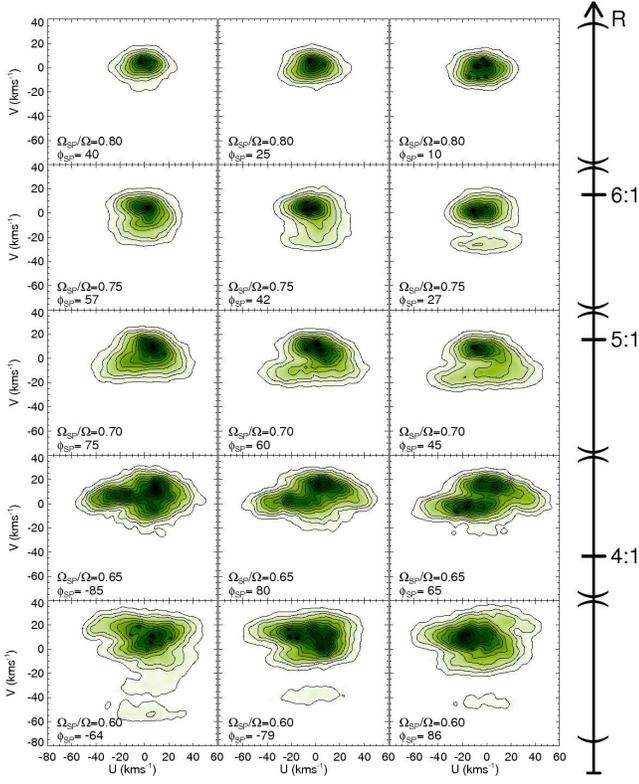} 
\caption{Same as Fig. \ref{f:speed18} but for locus 2. }  
\label{f:18locus2}
\end{figure}

 Now we compare the results with locus 1 and 2 (Fig. \ref{f:mw}). Fig. \ref{f:18locus2} shows the same case as in Fig. \ref{f:speed18} but using locus 2. The small difference between the pitch angles of these loci ($i=15.5\degg$ and $i=12.8\degg$, respectively), produces essentially the same kinematic structures when we compare regions with the same ratio $\Omega_{sp}/\Omega$ and relative spiral phase $\phi_{sp}$. Few differences are observed such as the group at low $V$ for left panel of the last row in Fig. \ref{f:18locus2} that does not appear in the equivalent relative position for locus 1 (approximately middle panel of last row of Fig. \ref{f:speed18}). These subtle differences are due to the tightness of the spiral arms for locus 2. 

Most of the 15 regions considered here correspond to interarm regions. When we analyse the velocity distributions in different regions of the Galactic disc, we see that the induced kinematic substructure increases near and on the spiral arms (Antoja et al., in prep.). For locus 2, we find more substructure than for locus 1 for the TWA because for this locus some of the regions are closer to the spiral arms. By contrast, for the \name model and the same pattern speeds, rich substructure is found for both locus. 

\begin{figure}
\centering
\includegraphics[width=0.35\textheight]{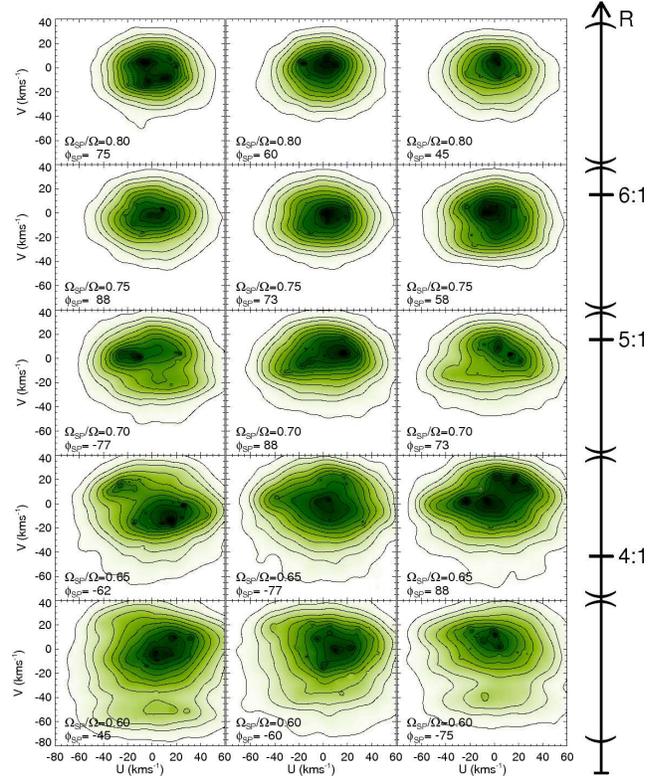} 
\caption{Same as Fig. \ref{f:speed18} but for the initial conditions IC2.}  
\label{f:18IC2}
\end{figure}

As expected the hotter population defined by the IC2 initial conditions (Section \ref{ic}) does not respond so strongly to the spiral perturbation. The main groups are still observed as the only changes are in the way each structure is populated but not in the orbital structure. The signal of some fine details slightly fades or appears less defined for the hotter case. We show an example in  Fig. \ref{f:18IC2}. Notice that the bi-modality in the first panel of the first row in Fig. \ref{f:speed18} that does not show up for IC2. Other structures manifested as separated groups are now stickied with particles filling the gaps in between.

\section{Strong kinematic substructure and resonances}\label{maximum}

Changing the pattern speed $\Omega_{sp}$ of the spiral arms corresponds to changing the position of the spiral resonances with respect to the regions under study. These resonances can be, as shown in other studies \citep{dehnen00,quillen05}, a major influence on the orbital structure. A detailed orbital analysis would be needed to determine the exact influence of each of the resonance in sculpting the velocity plane. Here, we examine the effects of the resonance proximity to the 15 regions of study. In the right part of e.g. Figs. \ref{f:speed18}, \ref{f:speed15} and \ref{f:speed20} we show a scale in galactocentric radius (configuration space) where we indicate the approximate radius of the main resonances. We also show in this scale the limits of the regions corresponding to the 3 adjacent velocity distributions. 

We determine that clear and rich kinematic substructure is seen near solar azimuth for the \name model for pattern speeds from 0.5 to 0.75 times the angular rotation rate. With a local circular frequency of $25.8\kmskpc$, this is for pattern speeds from 13 to 19.5 $\kmskpc$ at solar radius. This range of pattern speeds corresponds to being near the inner 3:1, 4:1 and 5:1 resonances. 

We also see that for pattern speed ratio of 0.76--0.85 ($\Omega_{sp}$ approximately between $19.5$ and $22\kmskpc$) we detect minor groups, that is, few and close to each other, or only small deformations of a unique clump centred in the \UVplane (e.g. second and third rows in Fig. \ref{f:speed20}). In these cases, the regions are located especially near the 6:1 inner resonance or high order $m$:1 resonances (with $m\ga6$). Therefore, we attribute weaker effects to these resonances. For pattern speed ratio approximately between 0.86 and 1.2 ($\Omega_{sp}$ between $22$ and $31\kmskpc$) we observe no substructure at the considered azimuths (e.g. upper row in Fig. \ref{f:speed20}). This is near corotation and high order resonances (inner and outer $m$:1 resonances with $m>6$), concluding that these resonances have no effect on the velocity plane at solar azimuths and for the maximum spiral arm density contrast. \footnote{Note that important substructure is also observed below the ratio 0.5 but is not considered as it is out of the established observational range. We also see that some substructure or deformation of a unique group for higher patten speeds ratios of 1.2--1.3. This is for pattern speeds from 31 to 33.5 $\kmskpc$, which is also out of the MW range. In this latter case, the effects are due to the approaching main outer resonances 4:1, 5:1 and 6:1.}

\begin{figure}
\centering
\includegraphics[width=0.35\textheight]{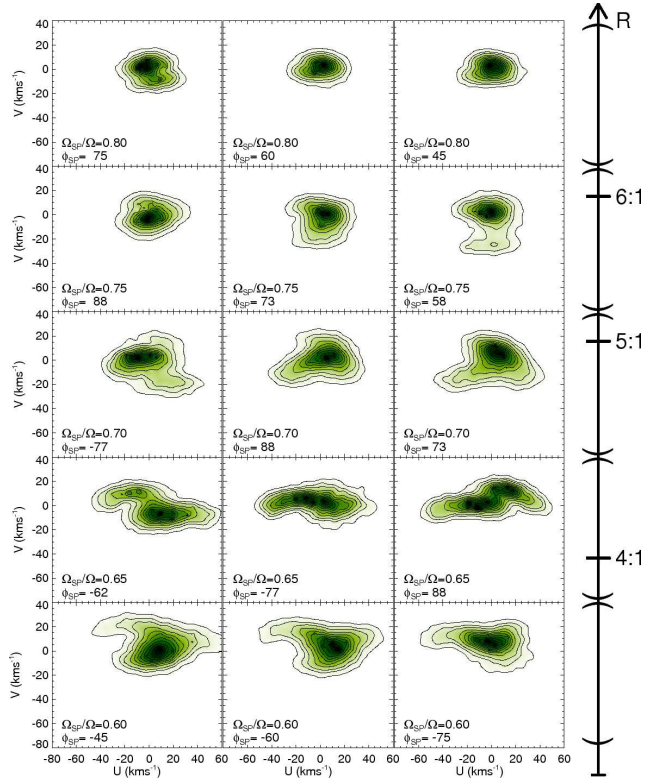} 
\caption{Same as Fig. \ref{f:speed18} but for the minimum spiral density contrast.}  
\label{f:18lowmass}
\end{figure}

 If we compare the results for the maximum density contrast with the lower spiral mass limit, the velocity distributions are similar but we see in most cases less populated or/and smaller kinematic groups for the low density case. As an example we examine in Fig. \ref{f:18lowmass} the same case as in Fig. \ref{f:speed18} but using the lower observational limit for the density contrast. Notice how the structures are essentially the same but the groups in the extremes of the \UVplane (far from the $(0,0)\kms$ point) disappear for the weaker spiral arms. For instance, the groups at low $V$ of last row of Fig. \ref{f:speed18} are not populated in Fig. \ref{f:18lowmass}. We observe this trend also for other pattern speeds. The range of pattern speeds where we find significant kinematic substructure is reduced with respect to the maximum density contrast case (0.5--0.75) and it now approximately 0.58--0.75. Thus being near the 3:1 inner resonances has lower influence for weaker arms. 

By examining more simulations, we see that the TWA gives less substructure than the \name model, given the same density contrast. It also gives clear and rich kinematic groups near solar azimuth for a smaller range of pattern speeds. Specifically, for the maximum density contrast, this range is 0.60--0.73 (pattern speed approximately from 15.5 to 18.8 $\kmskpc$). This is for regions close to the 4:1 and 5:1 inner resonances. For smaller pattern speeds, only minor substructures or small deformations of the velocity distributions are observed, contrary to the \name model. For larger pattern speeds, the induced kinematic groups disappear abruptly for the TWA, but progressively for the \name model. For a fixed pattern speed, all these lead to a smaller range of radius where substructure can be seen for the TWA. For pattern speed ratios larger than 0.8 (near corotation and high order resonances), we observe the same behaviour as in the \name simulations. The comparison between both models for the minimum density contrast is more drastic. The TWA model induces clear substructure only for ratios around 0.65--0.7 (particularly close to the 4:1 resonance). For the rest of the pattern speeds, almost no substructure is induced at the considered azimuths.

Previous results referring to the pattern speeds that give important or minor substructure do not depend substantially on the initial velocity dispersion or on the use of locus 1 and 2.

\section{Transient versus stationary effects}\label{time}

\begin{figure*}
\centering
\includegraphics[width=0.16\textwidth]{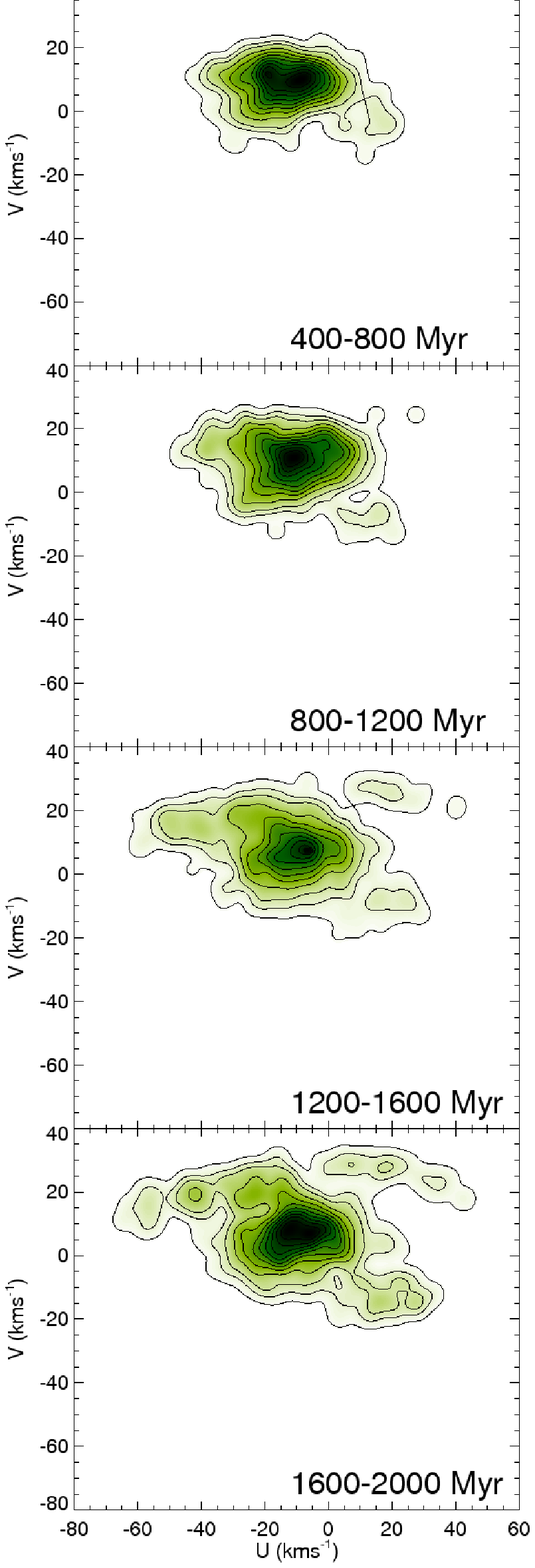}
\includegraphics[width=0.16\textwidth]{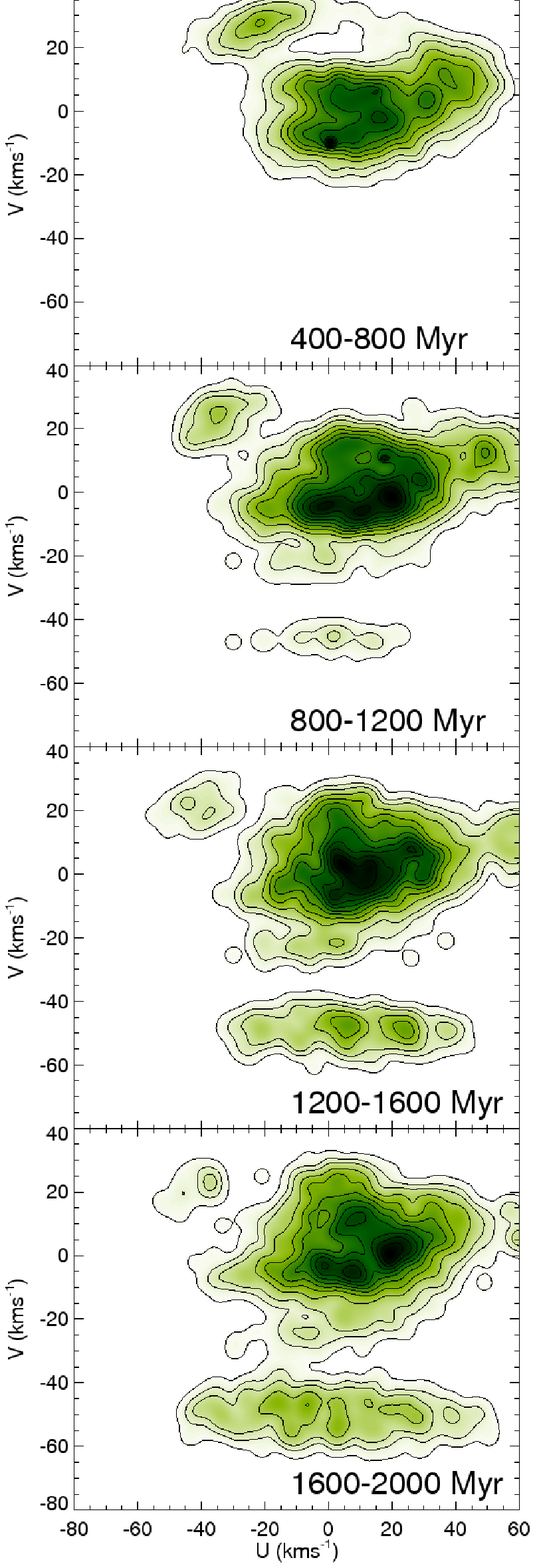}
\includegraphics[width=0.16\textwidth]{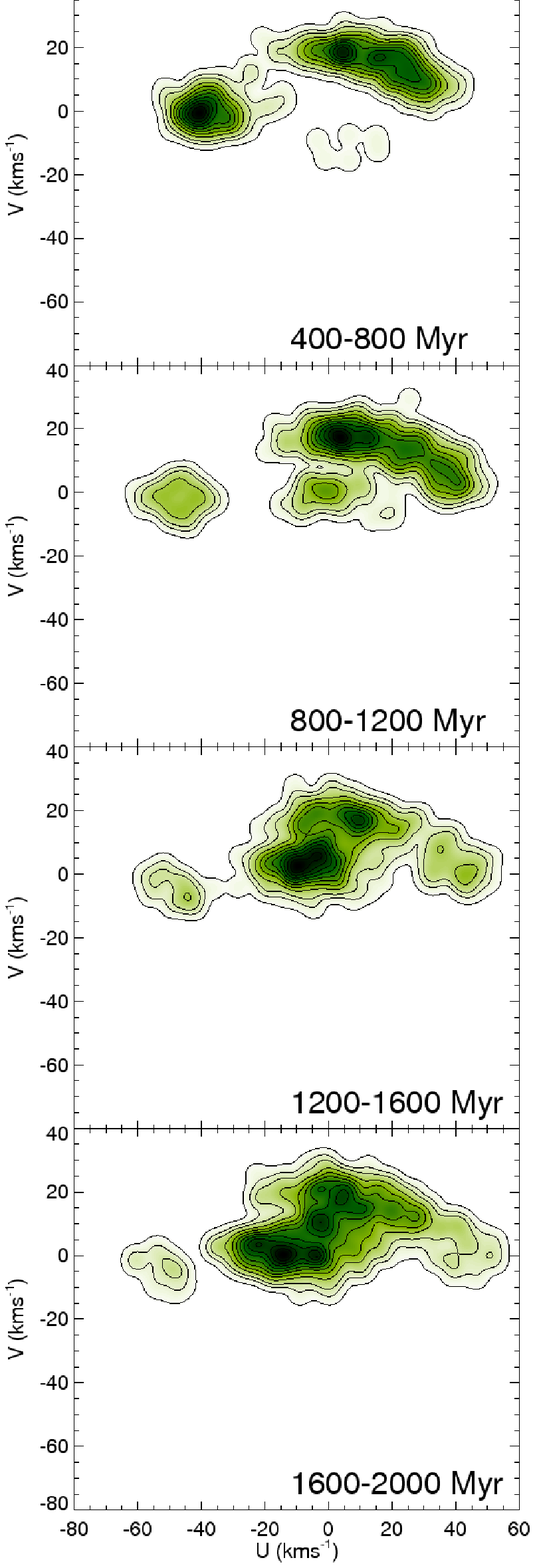}
\includegraphics[width=0.16\textwidth]{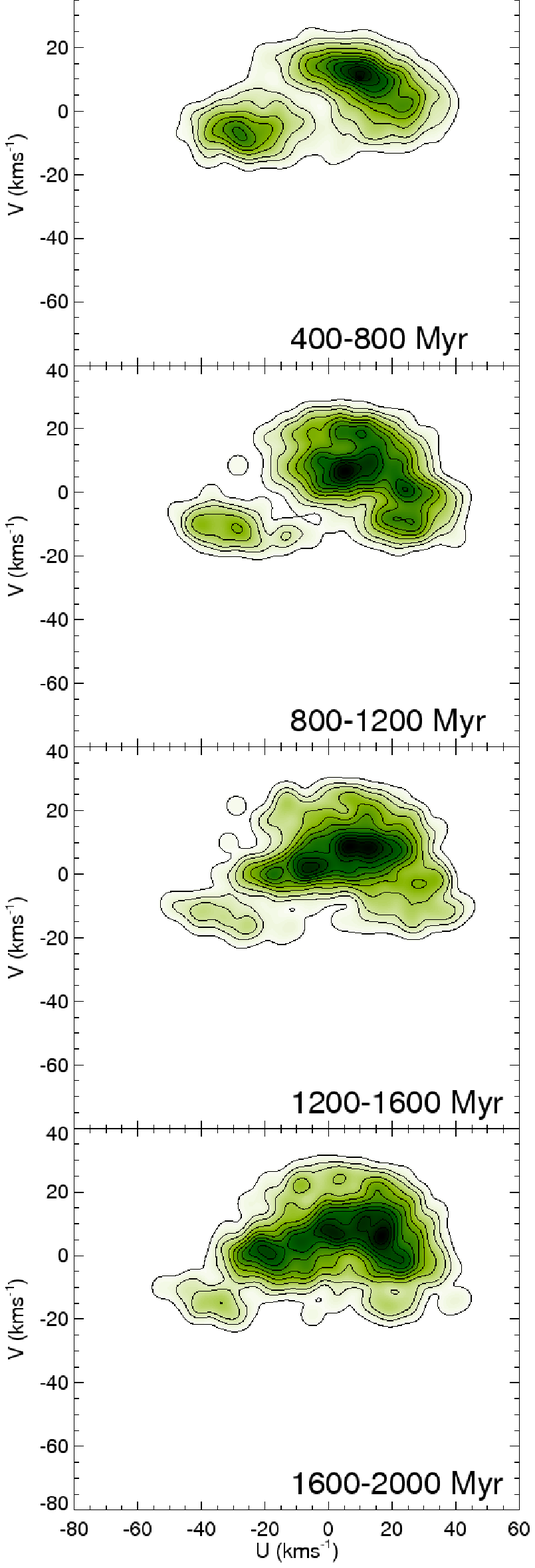}
\includegraphics[width=0.16\textwidth]{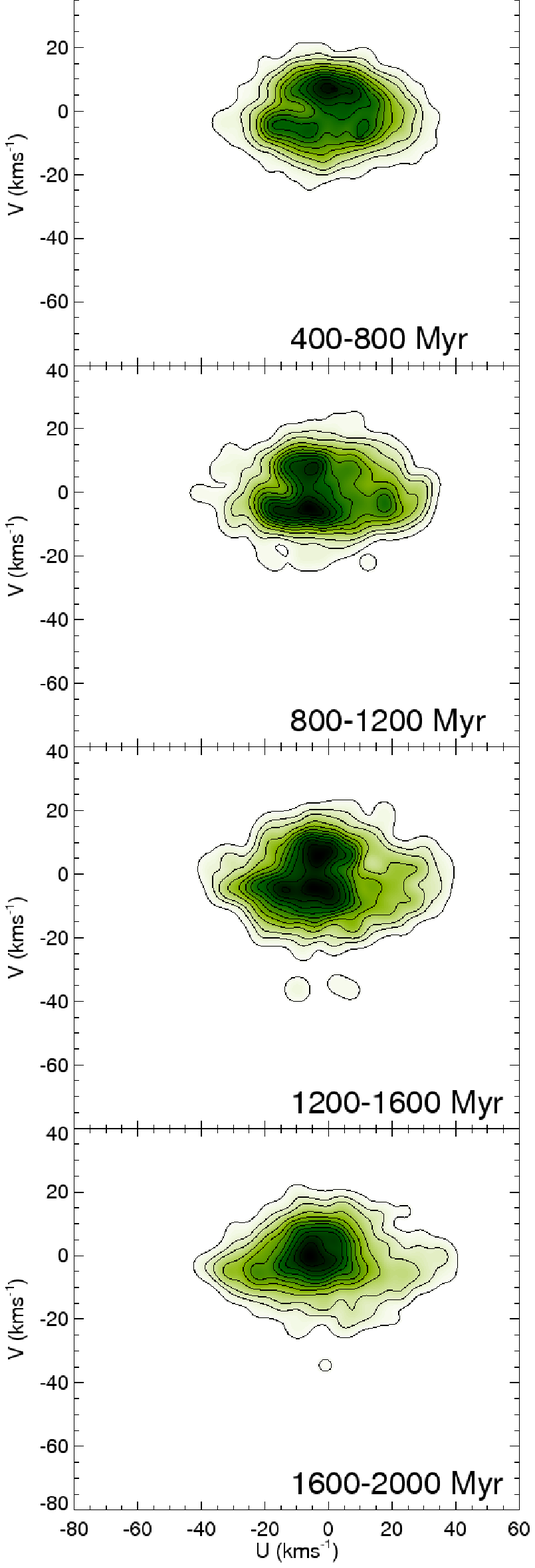}
\caption{$U$--$V$ velocity distributions as a function of integration time for several pattern speeds or ratios $\Omega_{sp}/\Omega$ and different $\phi_{sp}$ (indicated in the first panel of each column) for the \name model with locus 1, maximum density contrast and IC1.}  
\label{f:timePERLAS}
\end{figure*}

A fundamental question is if spirals are long or short-lived features. The processes that induce spiral arms play a key role in this \citep{sellwood10}. With the aim to see whether the velocity distributions can show imprints of the spiral lifetime, in this section we study the dependence of the structures on integration time. In our simulations, we have not modelled the progressive appearance of the spiral arms but they are included all time. In fact, some experiments showed that, given the same initial conditions, the final orbital structure does not change significantly with the adiabatic introduction of the non-axisymmetric component (Pichardo, private communication) or with different growth times \citep{minchev10}. In Fig. \ref{f:timePERLAS}, along columns, we show the time evolution of the \UVplane for 5 different representative cases (the pattern speed and relative spiral phase are indicated in the first panel of each column). We show the results for time integration intervals of 400 $\Myr$ from 0 to 2$\Gyr$, which is approximately one revolution period of the spiral arms. In these panels we see more separated groups compared to previous plots and, in general, the structures become separated with time. Structures in previous plots looked more continuous because they were composed by particles that had been integrated for a mix of times (between 0 and 2$\Gyr$).


The \UVplane of these simulations changes with time. This is due to the ongoing phase-mixing in response to spiral arm potential. We see that the time of appearance is different for each group, ranging from 0 to $\sim1200\Myr$. For instance, we observe two conspicuous groups in the first panel of second column (integration time smaller than 400 $\Myr$). Others appear later, as the branch at $V\sim-45$ for the period 800--1200 $\Myr$ in the model of second column. We observe how the structures at low $V$ or the ones at the extreme parts of the \UVplane require more integration time, i.e. a larger spiral lifetime. This is because they correspond to more eccentric orbits and larger radial excursions that take more time to reach the current region. We have also found that the time of appearance does not depend strongly on the density contrast of the spiral arms. On the other hand, a hotter disc (IC2) populates some structures at earlier times, specially at low $V$.

We see also that most structures change shape, size and position in the \UVplane with time. Some are formed in the central parts of the \UVplane and become progressively detached with time. Notice for example how the left structure in the third column shifts to negative more $U$. See also the branch at low $V$ in the second column which becomes more populated and more extended in the $U$ direction for large integration times. In general after 1200 $\Myr$ of integration time (3--4 revolutions of the spiral arms), the \UVplane becomes stationary; for this and larger times all the groups have already appeared and preserve the same size, shape and position in the kinematic space. Similarly, we saw that the global density structure became stationary after integration times between 400 and 1200 $\Myr$ (Section \ref{response}).

\section{Is it possible to constrain spiral arm parameters using local stellar kinematics?}\label{constrain}
In previous sections we have explored the effects of the spiral properties on the local velocity distribution. One would like to know which model parameter combination fits better the observed kinematic groups in the local \UVplanef. Here we discuss whether this constrain is currently possible.

\subsection{Degeneracy}

\begin{figure}
\centering
\includegraphics[width=0.23\textwidth]{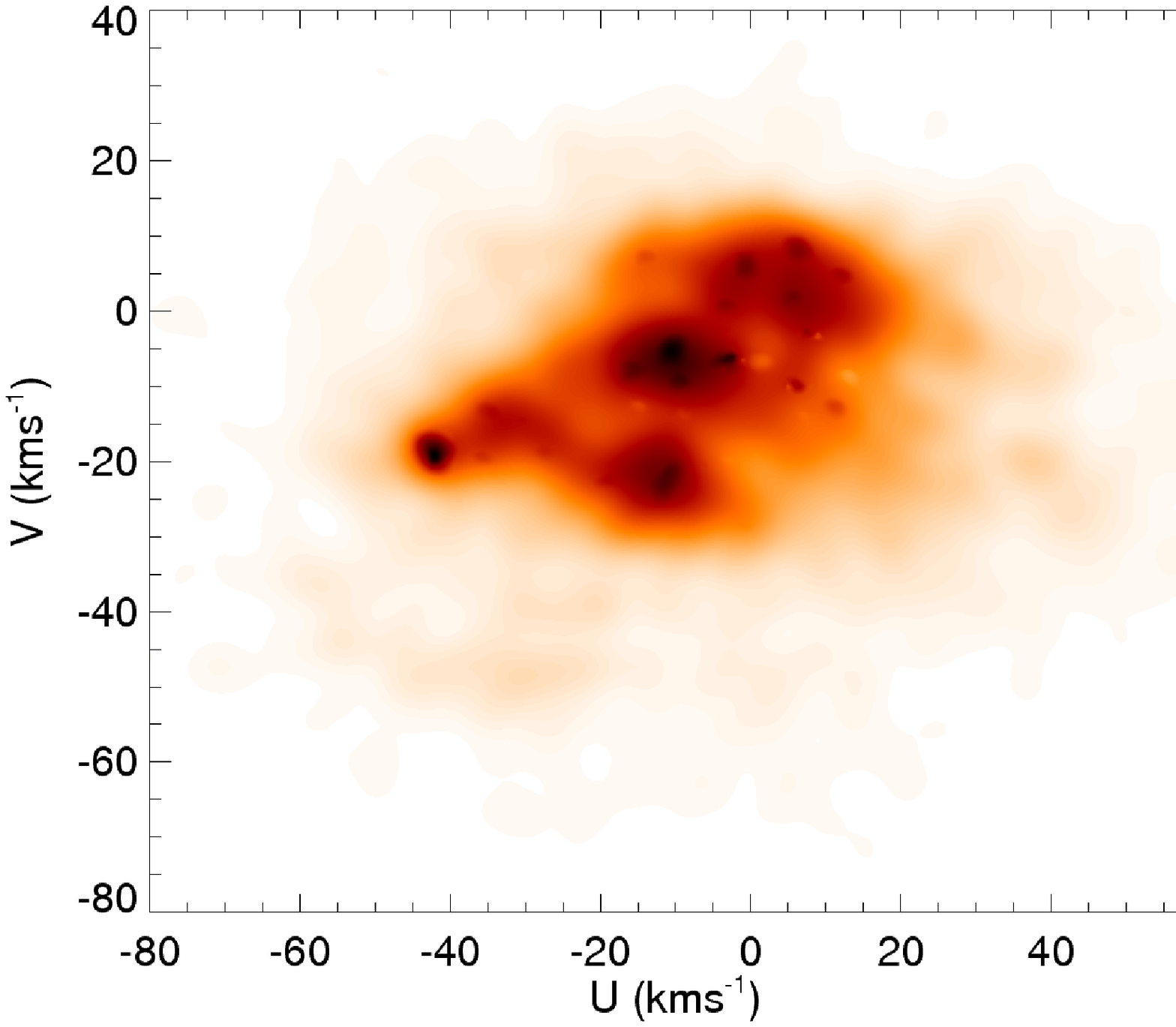}
\includegraphics[width=0.23\textwidth]{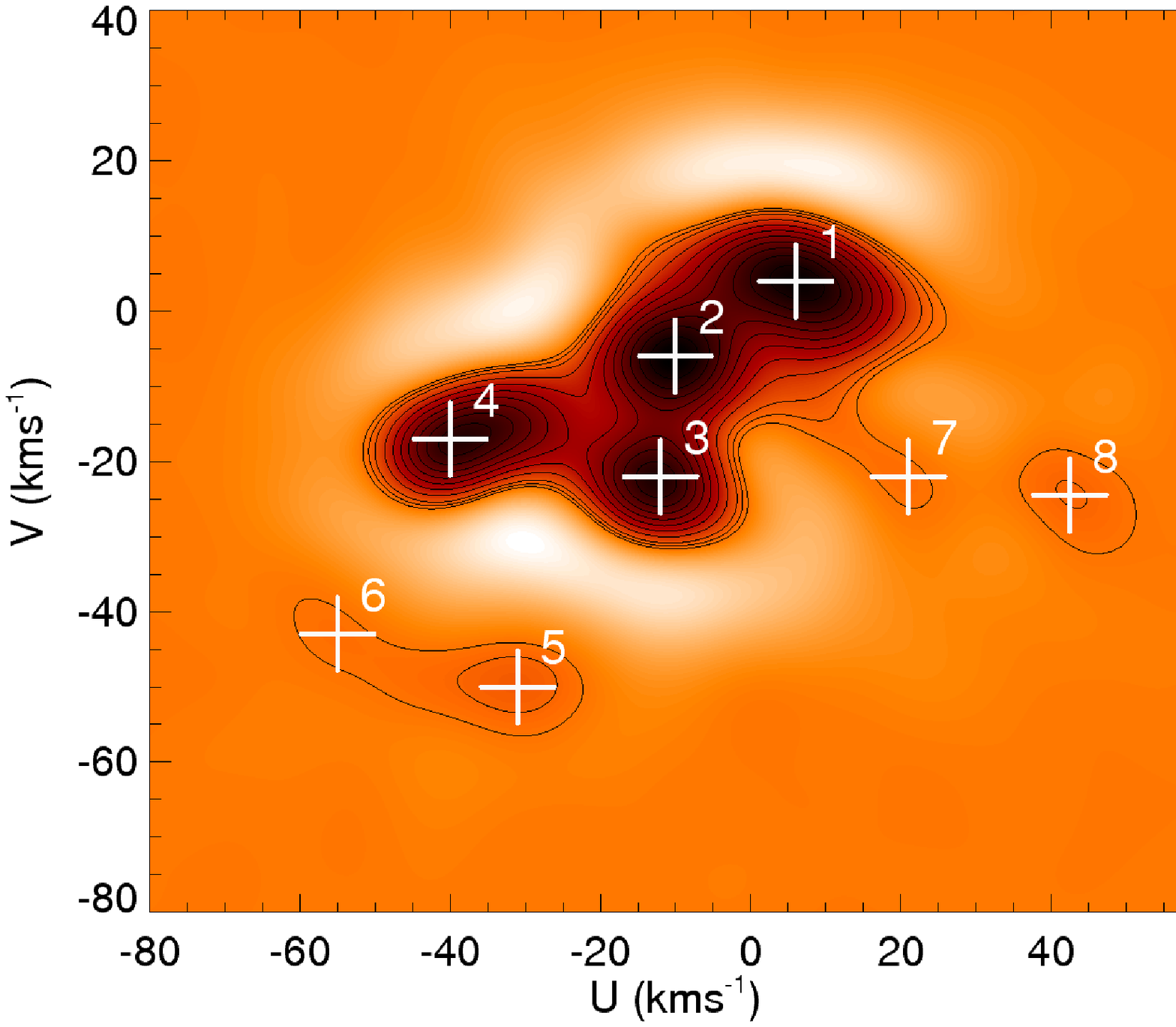}
\caption{Left: Observed heliocentric $U$--$V$ velocity distribution of the solar neighbourhood (data from \citealt{antoja08}). Right: Wavelet transform for a scale of $\sim10\kms$ of the same distribution.}  
\label{f:observed}
\end{figure}

The observed heliocentric velocity distribution of the solar neighbourhood is shown in Fig. \ref{f:observed} (left panel). This is obtained from the analysis in \citet{antoja08} but limited to stars with good velocity determinations (errors in all components $U, V, W$ $\leq 2\kms$). We have applied the Wavelet transform (see \citealt{antoja08}) to this distribution with a scale of $\sim10\kms$ (right panel of Fig. \ref{f:observed}), which highlights the kinematic structures of this size. The main groups of this distribution are Sirius (1), Coma Berenices (2), Pleiades (3) and Hyades (4),  groups 5 and 6 that account for the elongation of Hercules, and finally groups 7 and 8. To obtain velocities with respect to the LSR, we consider a solar motion of $(U_\sun,\ V_\sun,\ W_\sun)=(11,12, 7) \kms$ (Schoenrich, Binney \& Dehnen 2010). 
 Error bars of $5\kms$ in this plot account for the uncertainty in the definition of the exact group velocity.

\begin{figure}
\centering
\includegraphics[width=0.15\textwidth]{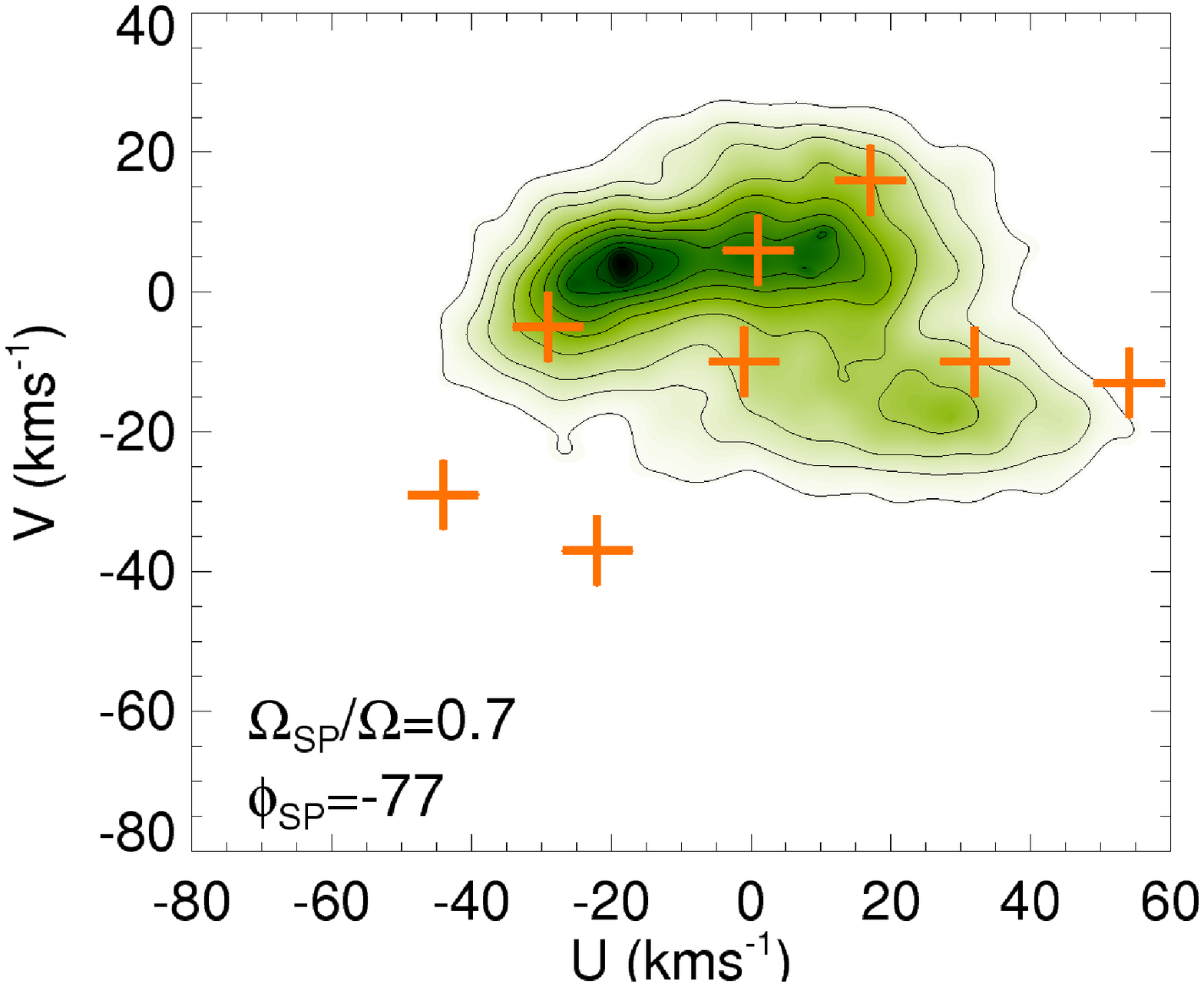}
\includegraphics[width=0.15\textwidth]{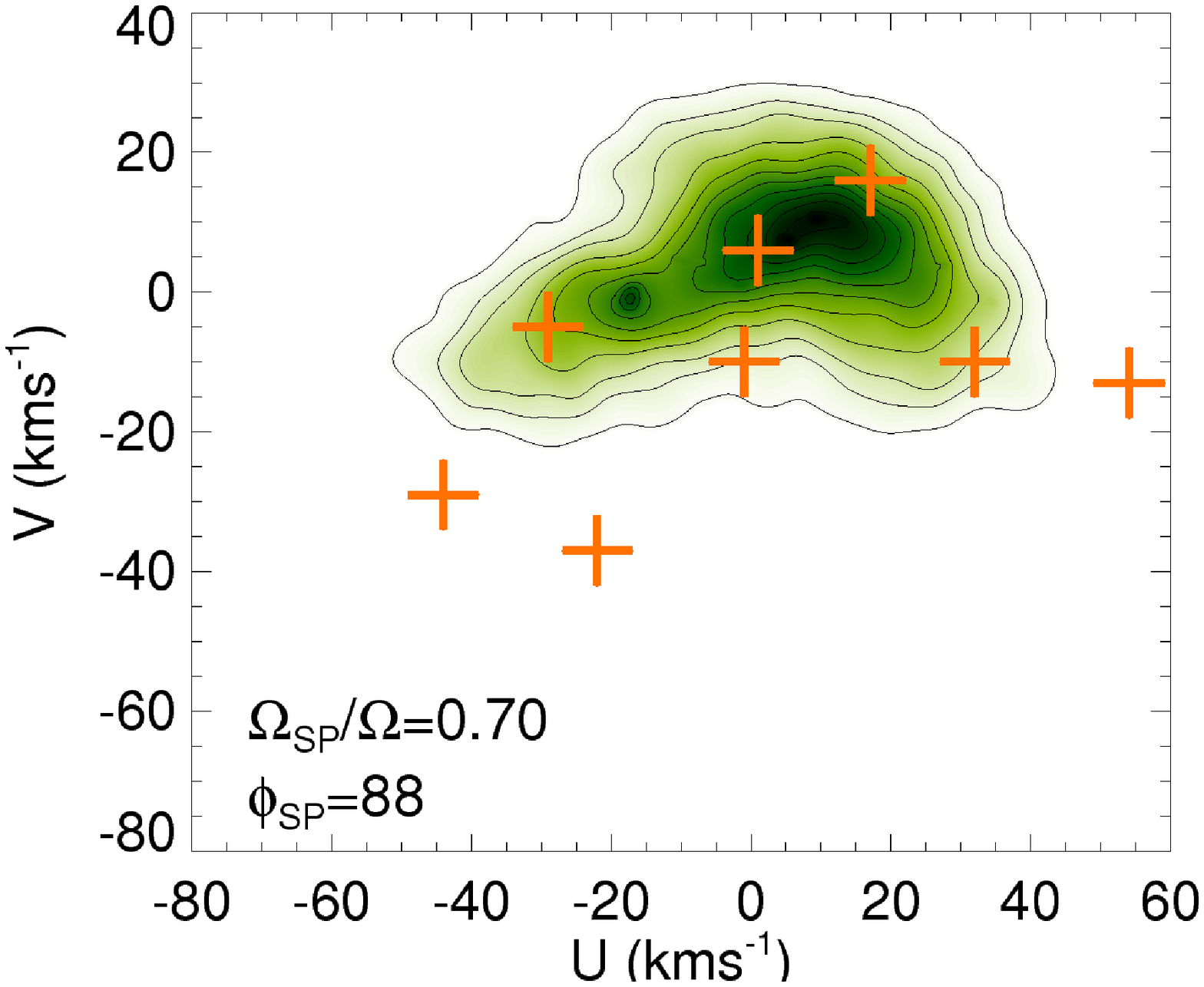}
\includegraphics[width=0.15\textwidth]{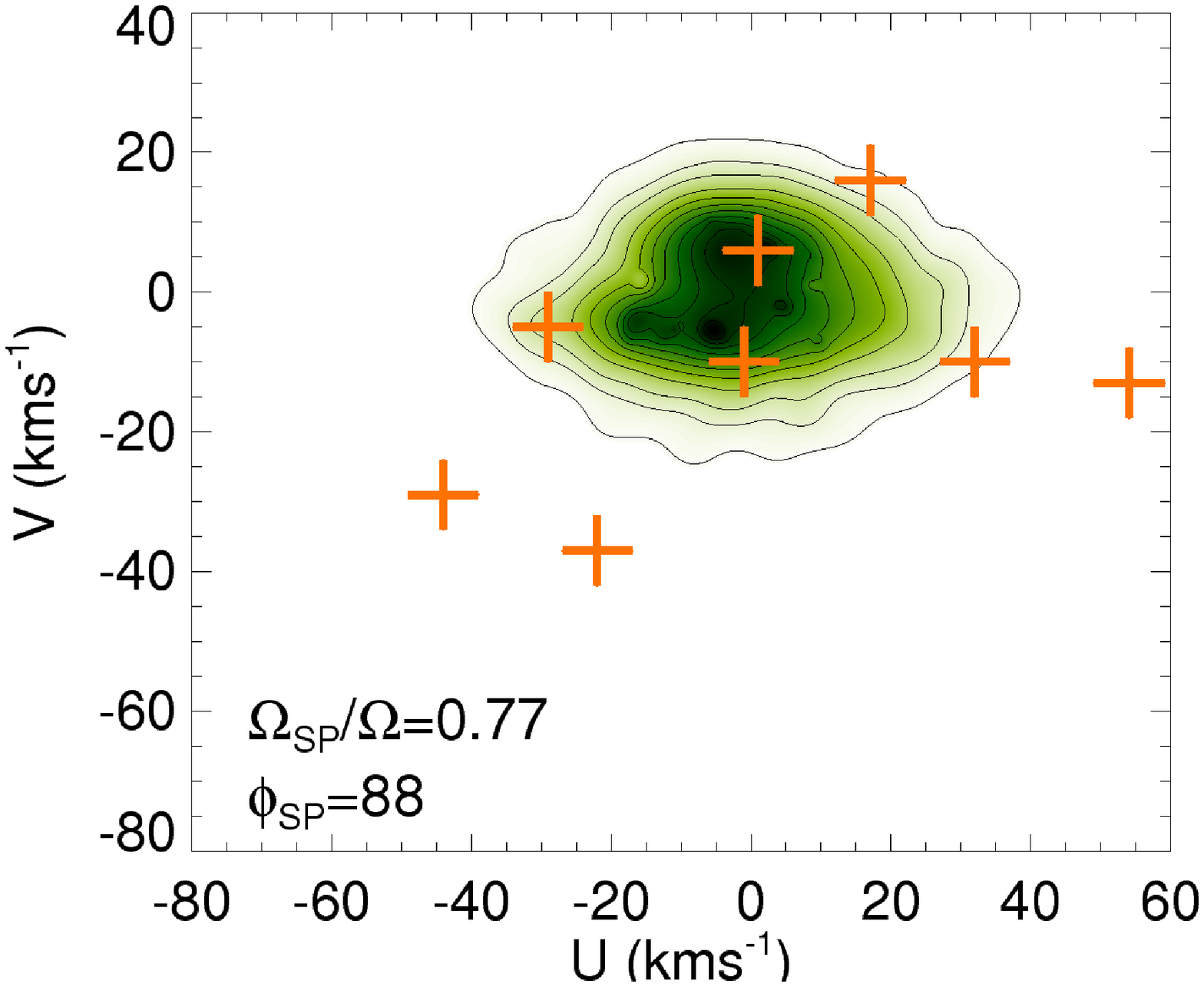}

\includegraphics[width=0.15\textwidth]{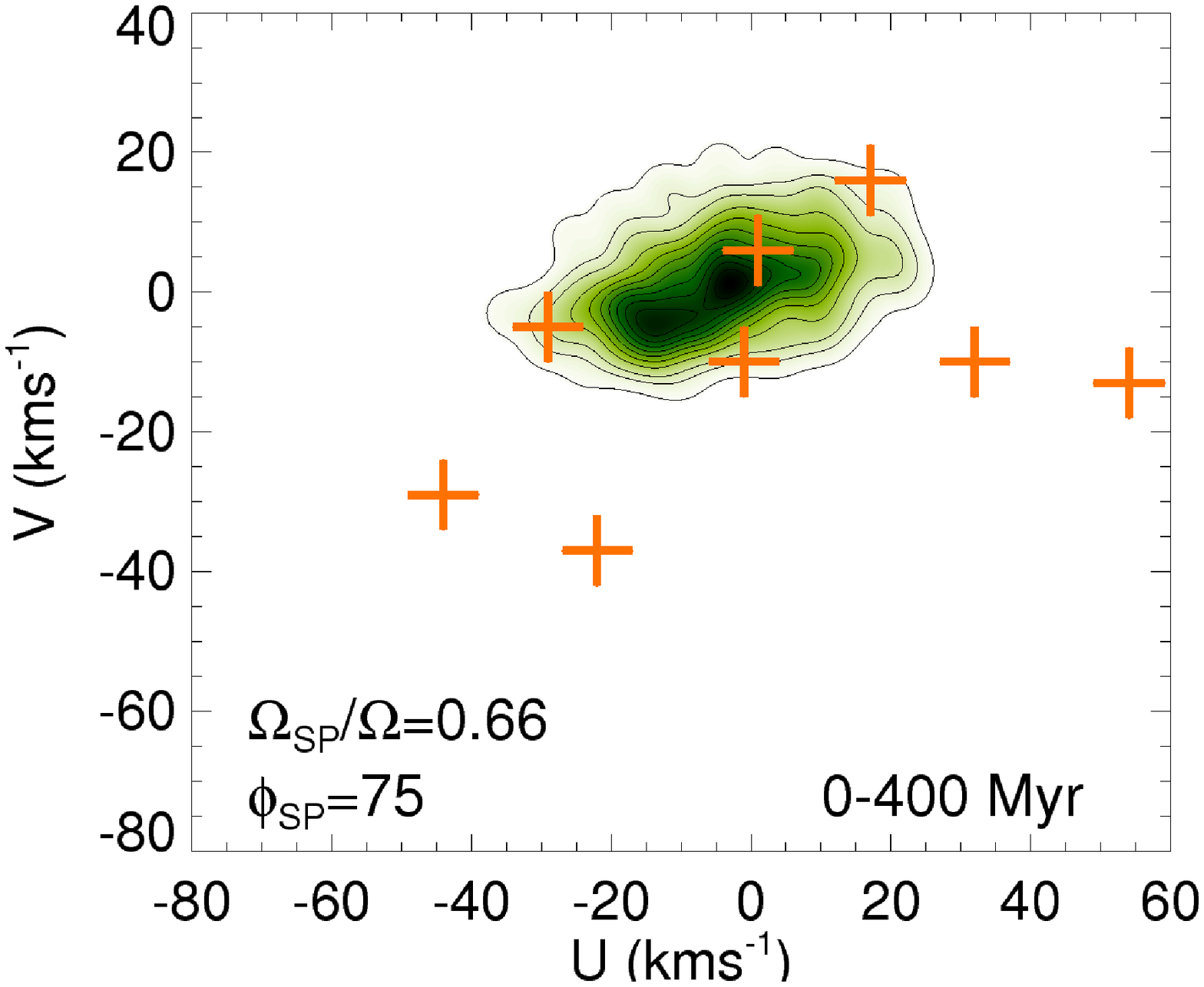}
\includegraphics[width=0.15\textwidth]{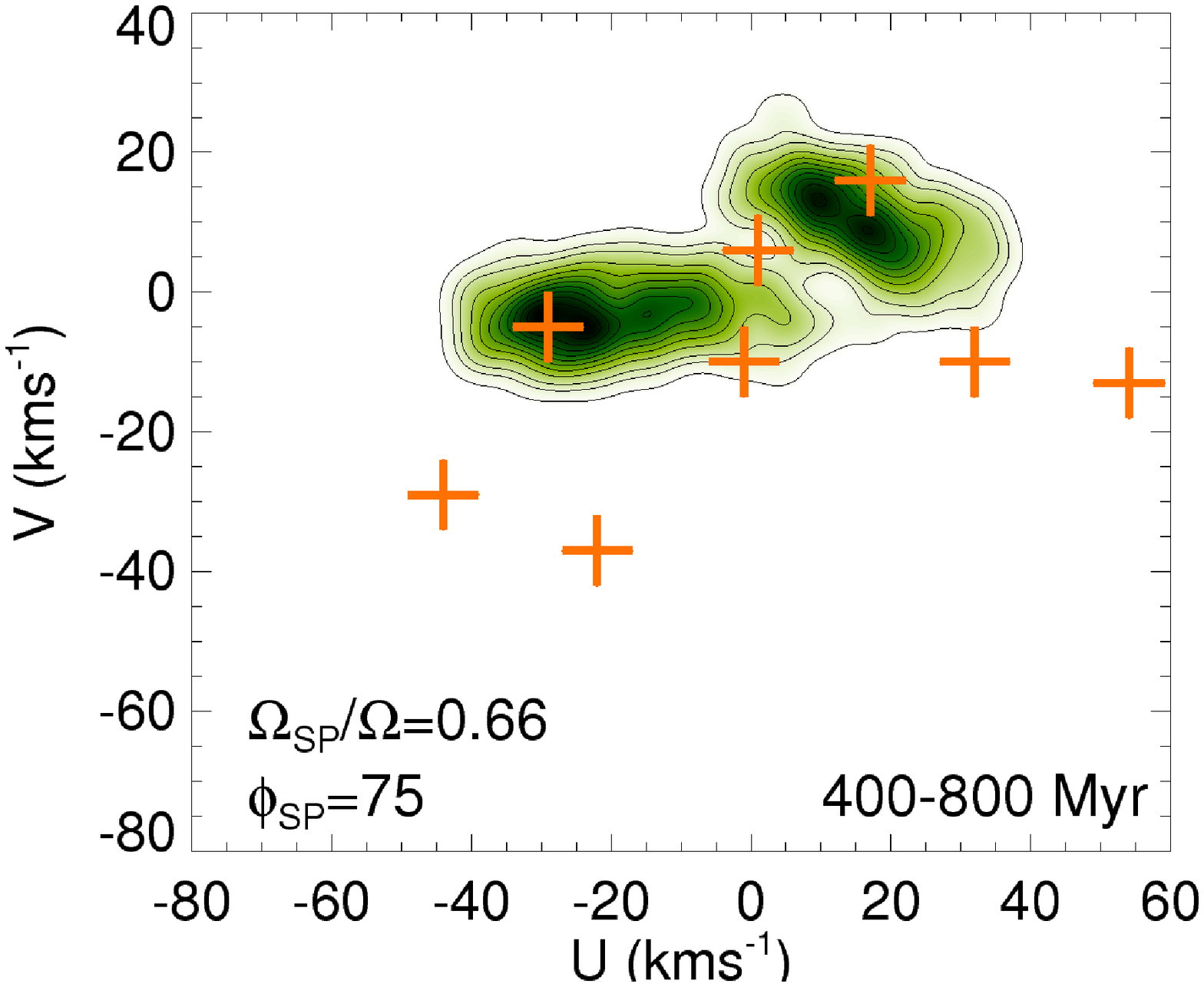}
\includegraphics[width=0.15\textwidth]{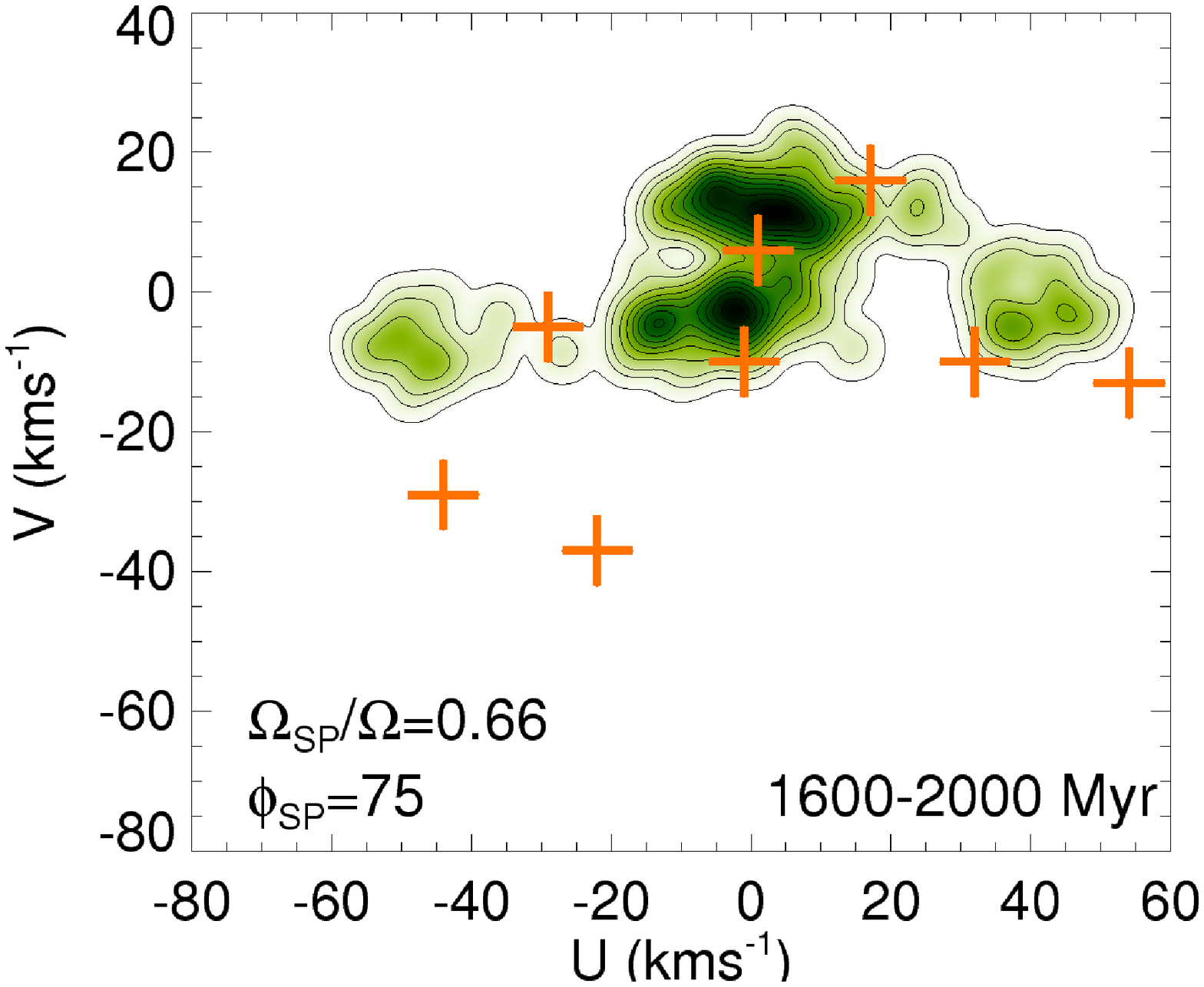}
\caption{$U$--$V$ velocity distributions for several models for the \name model with locus 1, maximum density contrast and IC1. Panels in the bottom show examples of the same model with different integration times. Orange points with error bars indicate the approximate positions of the local observed groups.}  
\label{f:constrain}
\end{figure}

We find several models, all within the established range of evidence for the MW spiral arms, that fit some of the observed groups in the solar neighbourhood, without any preference between these different ``good'' models. We also find that one particular group can be induced by different parameter combinations. We show in Fig. \ref{f:constrain} several models with the \name spiral arms that lead to kinematic groups similar to some of the observed ones. First panel in the first row fits Coma Berenices and presents elongations towards Hyades and groups 7 and 8. Second panel corresponds to the same pattern speed but a slightly different spiral phase and reproduces better Hyades and Sirius, showing also a small elongation towards group 7. Third panel seems to fit Coma Berenices and deviations towards Pleiades. Other good fits have been found for the TWA model. Panels in the bottom of Fig. \ref{f:constrain} correspond to the same model but different integration times. We see that especially for integration times from 400 to $800\Myr$ the model is able to reproduce Hyades and Sirius. This shows that the spiral lifetime could be also constrained. But the fact that other combinations of parameters have reproduced successfully these kinematic groups prevents us to adopt this as the best MW fitting. Notice that different authors have considered different observed groups and slightly different positions on the \UVplane due e.g. to different sample selections, or different solar motion determinations such as the one by \citealt{dehnen98b}). This ambiguity, however, does not change the conclusions of this section and, in fact, is an additional limitation to constrain the model parameters. 

This degeneracy is also noticed in the literature, where different models have led to the reproduction of the same group. For instance Coma Berenices have been related to the effects of the spiral arms \citep{quillen05} and to the effects of the bar \citep{minchev10}. Our simulations have shown that the parameter space is still large to obtain a unique best-fitting to the observed velocity distribution.

\subsection{Influence of other patterns and processes}\label{bar}

Apart from the mentioned current degeneracy, we should not expect an exact equivalence between the positions of the observed groups and the velocity distributions from the simplified currently available models. We must take into account that other processes may have influenced the local velocity distribution: the Galactic bar, external processes like past accretion events, or internal disc processes like star formation bursts or encounters with giant molecular clouds. 

\begin{figure}
\centering
\includegraphics[width=0.35\textheight]{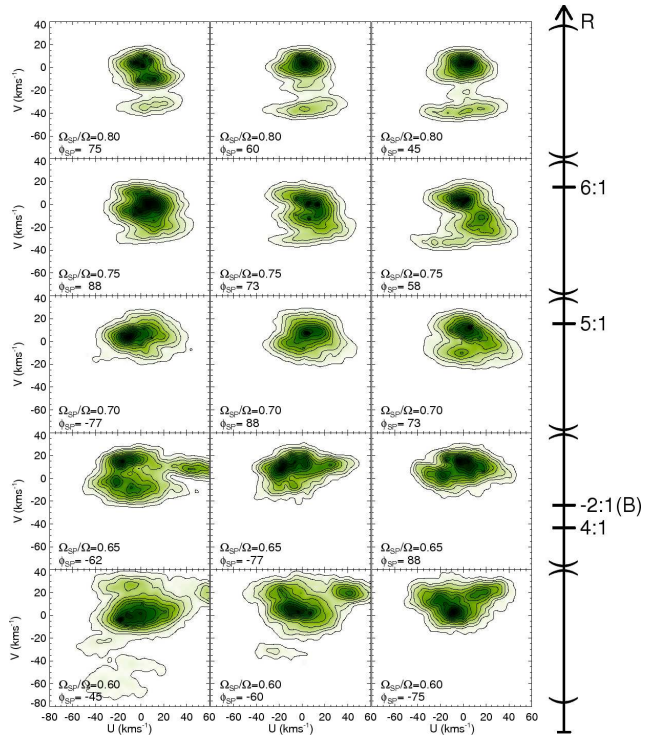} 
\caption{Same as Fig. \ref{f:speed18} but for a simulation including the potential of the Galactic bar with a pattern speed of $48\kmskpc$. The outer Lindlbad resonance of the bar is indicated as -2:1 (B) in the right scale.}  
\label{f:18spiralbar}
\end{figure}

The Galactic bar is believed to influence considerably the local velocity distribution \citep{dehnen00}. One may wonder how the velocity distributions presented in this study change under the combined action of spiral arms and bar. Several simulations have proved us that in such a combined case, the spiral arm imprints are still distinguishable. Fig. \ref{f:18spiralbar} are velocity distributions of the same simulation as in Fig. \ref{f:speed18}, but now including also a Galactic bar. The bar is modelled following MW observational constraints, as explained in \citet{antoja09}, and has a pattern speed of $48\kmskpc$ and present inclination with respect to the line Sun -- Galactic Centre of $20 \degg$. 
The total central mass of the model (bulge+bar) is $1.4\times 10^{10}\Msun$, approximately as estimated by \citet{dwek95}. Of this mass, 70\% belongs to the bar whereas the rest corresponds to the bulge, with a similar proportion as the best model in \citet{weiner99}. This mass gives maximums for the bar parameters $q_r$ and $q_t$ (defined in Section \ref{force}) of 0.25 and 0.37, respectively. These values are larger than for the spiral arms but are achieved at inner radii. In general, we see that in the inner regions of the disk (up to 3 or 4 $\kpc$) the bar dominates in strength but at solar radius of $8.5\kpc$ the strength values are comparable ($q_r$) or the spiral arms are stronger ($q_t$). 


As we see in Fig. \ref{f:18spiralbar}, the vast majority of structures created by the spiral arms are still seen in the combined case. In general, we see that the bar induces groups specially at the low part of the \UVplanef. For instance, we have check that the new group at $V\sim-40\kms$ in the first row appears also in the simulations with only bar. Whereas the central region of these 3 panels is rather similar to the spiral alone case. We find few structures that are not induced now (e.g. structure at low $V$ of the third panel in the last row), structures that have changed their shape (third panel in the fourth row) or that have been shifted to a slightly different position of the kinematic plane (structure at low $V$ of the first panel in the last row). The more severe differences are seen in the resonance overlap case (fourth row). This is expected as, according to \citet{quillen03}, this overlap can induce widespread chaos. Also \citet{chakrabarty07} claimed that the potential parameter constraint is not possible in these regions. But notice that we find that in nearby regions ($\sim600\pc$ in galactocentric radius) the individual spiral effects can still be identified. To conclude, using simulations with spiral-only model is a valid first step to understand the isolated effects of this non-axisymmetric component on the velocity distribution. 

It is particularly interesting that the arms (both models \name and TWA) can populate branch-like groups at around $V\la -40\kms$ (Section \ref{comparison}). The $V$ velocity of these branches is consistent with the $V$ velocity of the observed Hercules structure. This group has a $V$ heliocentric velocity between $-40$ and $-60\kms$ (\citealt{dehnen98}, see also Fig. \ref{f:observed}), which with the solar motion by \citet{schoenrich10}, corresponds to a $V$ velocity with respect to the LSR of $-30$ to $-50\kms$. Up to now this \UVplane region has been believed to be exclusively populated by the effects of the bar resonances. We show here and in \citet{antoja09} that spiral and unbarred models crowd structures at these negative $V$. Although its shape is not exactly equal to the Hercules branch, especially regarding its observed inclination in the \UVplane and average radial motion $\overline{U}<0$, we have just shown that the Galactic bar action can modify and shift the spiral-induced groups to more negative $U$. As a bar-only model is also able to induce a group similar to Hercules \citep{dehnen00,fux01,antoja09}, it is difficult to favour the bar or the combination of spiral arms and bar as the cause of this group.

The characteristics of the Galactic bar are still quite uncertain. For instance, determinations of the bar's angle with respect to the line Sun -- Galactic Centre range from $14\degg$ \citep{freudenreich98} to $40-45\degg$ \citep{hammersley00,benjamin05}. Moreover, the existence of one or two bars in the MW is currently being debated. It would be worth exploring in detail the kinematic effects of the combined spiral-bar allowed parameter space, although this is out of the scope of the present study. However, note that a different bar's orientation could produce some different kinematic groups, but will not change the conclusions of the section. This is not strictly true if we change the relative strength of the bar with respect to the arms or their pattern speeds. These may lead to kinematic distributions dominated by one of the patterns. Two extremes cases would be when one pattern exceeds the other one in strength or when its resonances are much closer to the considered regions. However, the ranges for the parameters of the spiral arms and bar in our models based on literature seem to indicate that the MW is far from these regimes. Simulations with other bar parameters, not presented in this study, have confirmed that the results of this section do not depend strongly on the parameters provided they are in the believed MW range.

We could also consider including other kind of patterns in the disc as proposed in the literature. For instance, several models have shown that galaxies might have more than one spiral arm mode (with different pattern speeds) at different radius or overlapping in radius (e.g. \citealt{rautiainen99}), or with pattern speed changing with radius (e.g. \citealt{dobbs10}). Some of the modes could be corotating with the bar at the bar's end, or coupled with the bar through their respective resonances. These possibilities are being explored for other galaxies (Meidt, Rand \& Merrifield 2009) but have hardly been addressed for the MW. These must be future points to take into account in our modelling.

\section{Summary and conclusions}\label{local}


Few studies have focused on the spiral arm effects on the local velocity distribution. Moreover, all of them modelled the spiral arm potential following the TWA. We have seen that, despite being uncertain, the observational evidences for the MW arms density contrast, and especially, the pitch angle, suggest that the assumptions for self-consistency of the TWA in the MW case are, at least, doubtfully satisfied. Here we have studied the spiral arm effects on the kinematics of the solar neighbourhood with the TWA but also with a different spiral arm model, the \name model. While in the TWA, the spiral arms are a small perturbative term of the potential, modelled as a cosine function, the \name spiral arms correspond to an independent 3D mass distribution, from which the gravitational forces are derived. We have seen that both radial and tangential forces of this model present more abrupt features and different disc locations for the minima, maxima and 0 points. 

Here we aimed to determine the conditions that favour the appearance of kinematic groups, such as the ones that we observe in the solar neighbourhood. To do this, first we have tuned these two models to the latest observational determinations for the properties of the MW spiral arms. For many of these properties only a certain range of evidence could be established. Next, we have performed test particle simulations with both models considering these ranges, different initial conditions and integration times. Our analysis indicates that:


\begin{enumerate}

\item A significant part of the allowed parameter space, especially for the \name model, favours the triggering of kinematic groups with different shapes, sizes and inclinations, such as the observed ones in the solar neighbourhood. This shows that it is feasible that some of the observed moving groups have a dynamical origin related with the spiral arms.

\item The velocity distribution is certainly sensitive to the relative spiral arm phase and, especially, to the pattern speed. Changes in the kinematic groups are significant if one moves only $\sim0.6\kpc$ in radius. But the \UVplane changes more slowly with azimuth and $\sim 2\kpc$ are needed in this direction to detect important differences. Due to this, the pattern speed could be better constrained using the observed kinematic groups (within an error of $\la 2\kmskpc$) than the relative spiral phase (with a precision of $\sim15\degg$). However, one would need to break the degeneracy mentioned below by other means such as reducing the free parameter space. 

\item For both models and for all density contrasts, within the observational MW range, the spiral arms induce strong imprints for pattern speeds around $17\kmskpc$. This corresponds to regions close to the 4:1 inner resonance.  No substructure at all is induced close to corotation or high order resonances ($m>6$), which corresponds to pattern speeds of 20.5 to 30 $\kmskpc$. 

\item Changes in spiral strength produce no significant differences in most cases, which makes difficult to constrain this parameter. Some groups in the extremes of the \UVplane are only populated for the maximum density contrast case, which could help to establish the spiral strength. The effects of a different pitch angle seem difficult to differentiate given the observational range for this parameter. 

\item The spiral arms induce groups in a large region of the \UVplanef, including the low $V$ part. For instance, the arms, and not only the bar, can crowd the region of the Hercules group. Previous studies associated the spiral arm influence mainly to the central parts of the \UVplanef, and the bar to low angular momentum structures, such as Hercules \citep{quillen05,dehnen00}. 

\item The kinematic groups that are dynamically induced by the spiral arms depend on integration time. The time of appearance is different for each group, ranging from 0 to 1200 $\Myr$. The structures at the extreme parts of the \UVplane require more integration time, i.e. a larger spiral lifetime. The early appearance of some groups demonstrates that even recent spiral arms ($< 400\Myr$) may produce strong kinematic structures. Structures also change shape, size and position in the \UVplane until a maximum time of $\sim1200\Myr$, when the \UVplane becomes stationary. 

\item Each of the structures seen in the \UVplane is composed by particles with a wide range of integration times and a different minimum time. This is encouraging as the study of the stellar ages in the observed moving groups also reveal similar characteristics. For instance, see evolution in the Sirius and Hyades groups particularly for stars younger than $2\Gyr$ (fig. 13 in \citealt{antoja08}) or the wide age distribution and different minimum age in a particular group (fig. 14 in \citealt{antoja08}). 

\item Models obtained from our simulations with the spiral arms that reproduce the local observed velocity distribution are degenerated. Groups such as the ones that are observed in the solar neighbourhood can be induced by different model parameter combinations. For instance, several models create groups such as Hyades, Pleiades or Sirius. 

\item In most of our simulations where both the spiral arms and the bar are included, individual imprints of the bar and the arms can still be identified in the final velocity distributions. This means that using simulations with spiral-only model is a valid first step to understand the isolated kinematic effects of this non-axisymmetric component.


\item The stellar response near solar positions of the TWA spiral arms and the \name model is significantly different, both in configuration space and kinematics. Both models are able to induce several kinematic groups but the velocity distributions and the groups are rather different. 

\item The \name model gives more substructure than the TWA given the same density contrast, and more substructure for a larger pattern speed range or, equivalently, a larger radius range. For the \name model and both the maximum and minimum density contrast, we see clear and rich kinematic substructure near solar azimuth for pattern speeds from 15 to 19.5 $\kmskpc$ (regions near the inner 3:1, 4:1 and 5:1 resonances). By contrast, this pattern speed range is smaller for the TWA model even in the high density contrast case. On the other hand, for the minimum density contrast, the TWA model induces some substructure only for a narrow range of pattern speeds (around 16.7--18 $\kmskpc$).

\end{enumerate}



To conclude, the spiral arms induce strong imprints on the velocity distributions at solar regions and that these are sensitive to some arm properties. These indicate that kinematics can be used as one of the constraints on the current uncertainty about spiral structure of our Galaxy, e.g., in the pattern speed, strength, orientation and lifetime. 

The local distribution function is significantly sensitive to the used model, even if they are adjusted to reproduce the same observational constraints for the spiral structure of the MW. This stresses the importance of the specific spiral arm gravitational potential modelling for the MW studies, as this can induce significant deviations on their orbital dynamics. It is promising that the kinematic structures can help to improve this modelling or to test whether the MW spiral arms are weak and tightly wound, following the TWA.

The constraint of the MW spiral structure properties, however, is presently not straightforward. This is due, first, to the mentioned degeneracy. Our simulations have shown that the parameters space is still too large to obtain a unique fitting to the observed velocity distribution. Second, the uncertainty in the solar motion propagates to the velocities of the observed groups, making the fitting between observations and simulations imprecise. The problem is complex as, precisely, the presence of kinematic substructure complicates the solar motion determination. Third, for this constraint we should take into account the other processes that have influenced and sculpted the real velocity distributions (other non-axisymmetries, external accretion effects, dispersion by giant molecular clouds,etc). As an example, we have identified effects due to the combination of bar and spiral arms. For instance, structures that appear centred in the $U$ axis in the spiral-only models are shifted to negative $U$ due to the combined action, inducing a group similar to the observed Hercules structure. Nevertheless, none of the existing current Galaxy models is complex enough to include all these processes at the same time. 


To break the degeneracy and use the kinematic groups to constrain the MW large scale structure, a smaller parameter space for the spiral properties and/or data from velocity distributions at different positions of the MW disc are needed. With the increase of knowledge on the MW structure and evolution, we will have to cope with a more complex scenario and a variety of processes that can play a role in the formation of moving groups. Future models will be more realistic but more complex to interpret. Besides, while now the spatial study of the observed moving groups is limited by the extension and precision of the current observational samples, this will soon change with the advent of data from new surveys (USNO, UCAC3, 2MASS, RAVE, SEGUE, PanStars, LAMOST, Gaia). Explorations of the spiral arm effects like the present study will help us to interpret more sophisticated future models as well as velocity distributions in nearby regions coming from surveys such as Gaia.

\section*{Acknowledgments}

We thank the referee, P. Rautiainen, for useful suggestions that helped improve this manuscript. We acknowledge funding support from the European Research Council under ERC-StG grant GALACTICA-24027, MICINN (Spanish Ministry of Science and Innovation) - FEDER through grant AYA2009-14648-C02-01 and CONSOLIDER CSD2007-00050. Some of the simulations were run at the HP CP 4000 cluster (KanBalam) in the DGSCA/UNAM.

\label{lastpage}

\end{document}